\documentclass[prx,amsmath,amssymb,aps,superscriptaddress,showkeys,floatfix]{revtex4-2}

\usepackage{graphicx}
\usepackage{bbm}
\usepackage{subcaption}
\usepackage{physics}
\usepackage{hyperref}
\hypersetup{hidelinks}

\newcommand{\kb}[1]{\ket{{#1}}\!\bra{{#1}}}

\newcommand{\be}{\begin{equation}}
\newcommand{\ee}{\end{equation}}
\newcommand{\bea}{\begin{eqnarray}}
\newcommand{\eea}{\end{eqnarray}}
\newcommand{\bes}{\begin{equation*}}
\newcommand{\ees}{\end{equation*}}
\newcommand{\beas}{\begin{eqnarray*}}
\newcommand{\eeas}{\end{eqnarray*}}
\newcommand{\bpm}{\begin{pmatrix}}
\newcommand{\epm}{\end{pmatrix}}

\newtheorem{theorem}{Theorem}

\newtheorem{proposition}[theorem]{Proposition}

\usepackage{float}
\usepackage{booktabs}

\begin{document}

\title{Unified Probabilistic Error Cancellation and Zero-Noise Extrapolation for Mid-Circuit Quantum Measurements}

\affiliation{Graduate School of Quantum Science and Technology, Korea Advanced Institute of Science and Technology (KAIST)}
\affiliation{School of Electrical Engineering, Korea Advanced Institute of Science and Technology (KAIST), 291 Daehak-ro, Yuseong-gu, Daejeon 34141, Republic of Korea}
\affiliation{Center for Superconducting Quantum Computing System, Korea Research Institute of Standards and Science (KRISS), 267 Gajeong-ro, Yuseong-gu, Daejeon 34113, Republic of Korea}

\author{Hyunseong Jang}
\email{hyunseong@kaist.ac.kr}
\affiliation{Graduate School of Quantum Science and Technology, Korea Advanced Institute of Science and Technology (KAIST)}
\affiliation{Center for Superconducting Quantum Computing System, Korea Research Institute of Standards and Science (KRISS), 267 Gajeong-ro, Yuseong-gu, Daejeon 34113, Republic of Korea}

\author{Hwan-Seop Yeo}
\affiliation{Center for Superconducting Quantum Computing System, Korea Research Institute of Standards and Science (KRISS), 267 Gajeong-ro, Yuseong-gu, Daejeon 34113, Republic of Korea}

\author{Jiman Choi}
\affiliation{Center for Superconducting Quantum Computing System, Korea Research Institute of Standards and Science (KRISS), 267 Gajeong-ro, Yuseong-gu, Daejeon 34113, Republic of Korea}

\author{Jiwan Song}
\affiliation{Center for Superconducting Quantum Computing System, Korea Research Institute of Standards and Science (KRISS), 267 Gajeong-ro, Yuseong-gu, Daejeon 34113, Republic of Korea}

\author{Yong-Ho Lee}
\affiliation{Center for Superconducting Quantum Computing System, Korea Research Institute of Standards and Science (KRISS), 267 Gajeong-ro, Yuseong-gu, Daejeon 34113, Republic of Korea}
\affiliation{Graduate School of Quantum Science and Technology, Korea Advanced Institute of Science and Technology (KAIST)}

\author{Joonwoo Bae}
\email{joonwoo.bae@kaist.ac.kr}
\affiliation{Graduate School of Quantum Science and Technology, Korea Advanced Institute of Science and Technology (KAIST)}
\affiliation{School of Electrical Engineering, Korea Advanced Institute of Science and Technology (KAIST), 291 Daehak-ro, Yuseong-gu, Daejeon 34141, Republic of Korea}

\begin{abstract}
Mid-circuit measurements enable adaptive quantum computation, but their errors affect both classical feedforward outcomes and the quantum states entering subsequent operations. Mitigating these errors therefore requires treating measurements as quantum instruments rather than correcting classical readout statistics alone. In this work, we develop a unified framework for probabilistic error cancellation (PEC) and zero-noise extrapolation (ZNE) of Pauli-tailored mid-circuit measurement instruments. We formulate an outcome-resolved stochastic noise model that incorporates errors in measurement outcomes and conditional post-measurement states, including their correlations. Using finite-group convolution, we derive quasiprobability coefficients for ideal instrument reconstruction and establish an affine relation between PEC reconstruction and controlled noise amplification. We further identify the conditions under which amplified measurement noise can be physically implemented through stochastic sampling of the same Pauli-dressed operations. We investigate the framework through numerical simulations of iterative quantum phase estimation and a proof-of-principle demonstration using a 20-qubit superconducting quantum processor developed at the Korea Research Institute of Standards and Science (KRISS), where two qubits are exploited to implement adaptive phase estimation with mid-circuit measurements, feedforward, and qubit reuse. Using experimentally characterized measurement errors, we implement both PEC and stochastic noise amplification for ZNE, obtaining phase-expectation point estimates closer to the ideal reference than conventional readout correction. Our work provides a systematic approach to instrument-level error mitigation in adaptive quantum circuits.
\end{abstract}

\maketitle

\section{Introduction}

\begin{figure*}[t]
\includegraphics[width=0.98\textwidth]{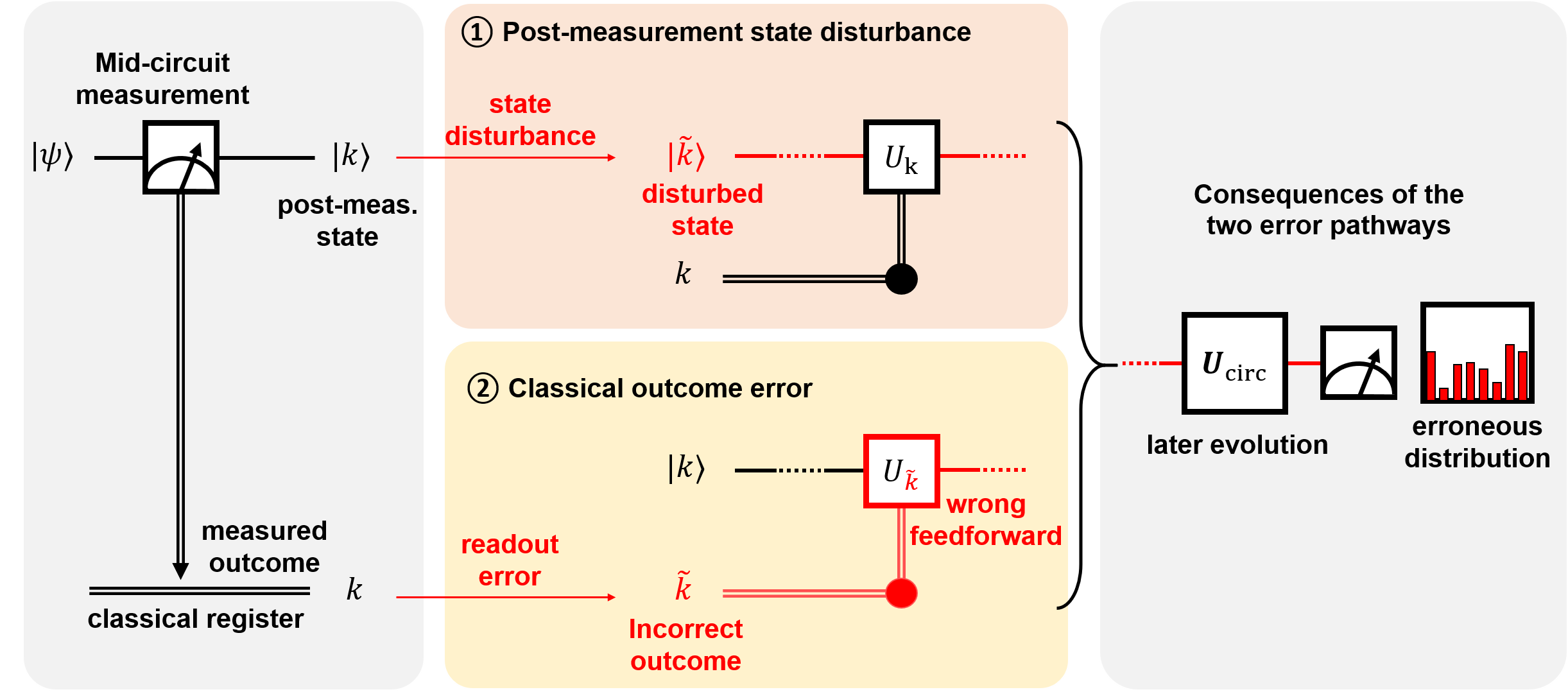}
\caption{Two error pathways in adaptive circuits are shown. A disturbed post-measurement state can propagate to conditional operations even when the logical record is correct (top), whereas an incorrect mid-circuit measurement record can select the wrong conditional gate (bottom). Both mechanisms can bias downstream observables.}
\label{fig:DC}
\end{figure*}

Mid-circuit measurements (MCMs) are essential building blocks of adaptive quantum computation, enabling measurement-conditioned feedforward, active reset, qubit reuse, and dynamic and fault-tolerant quantum circuits~\cite{PhysRevLett.133.150602,RevModPhys.87.307,548464,PhysRevLett.76.3228}. Unlike terminal measurements, which primarily determine the final classical output of a circuit, MCMs directly influence subsequent quantum operations. Their outcomes determine the classical control flow, while their post-measurement states serve as inputs to the remaining quantum evolution. Consequently, measurement errors can affect an adaptive circuit through two distinct mechanisms: an erroneous classical outcome can trigger faulty feedforward, while a disturbed conditional quantum state can propagate errors even when the recorded outcome is correct, see Fig. \ref{fig:DC}. Reliable adaptive quantum computation therefore requires mitigation strategies that account for both the classical measurement record and the resulting quantum state.

Conventional measurement-error mitigation typically relies on a calibrated assignment or confusion matrix to reconstruct ideal measurement-outcome probabilities~\cite{PhysRevA.103.042605,PRXQuantum.2.040326,PhysRevA.105.032620,Maciejewski2019,9142431}. Although such approaches can correct classical readout statistics, they do not generally account for errors in the conditional post-measurement quantum states. This limitation is particularly important in adaptive quantum circuits, where measurement outcomes determine subsequent feedforward operations and the corresponding quantum states undergo further evolution. Even when the ideal measurement-outcome distribution is recovered, unmitigated errors in the conditional quantum states can propagate through subsequent operations, leading to biased expectation values of final observables. Addressing these errors therefore requires treating an MCM as a quantum instrument, described by a collection of outcome-conditioned completely positive maps that jointly characterize the classical measurement record and the associated quantum-state transformation. This instrument-level perspective motivates error-mitigation strategies that reconstruct conditional quantum operations rather than correcting classical measurement statistics alone.

Among quantum error-mitigation techniques, probabilistic error cancellation (PEC) and zero-noise extrapolation (ZNE) provide two complementary approaches. PEC reconstructs ideal operations through signed quasiprobability combinations of experimentally accessible noisy operations, whereas ZNE estimates ideal expectation values by evaluating a circuit at different noise strengths~\cite{PhysRevLett.119.180509,PhysRevX.8.031027,PhysRevX.7.021050}. Recent studies have extended PEC to dynamic operations, including measurement errors and feedforward~\cite{PhysRevA.109.062617}, and explored measurement randomized compiling and quasiprobabilistic readout correction for mid-circuit feedback~\cite{PRXQuantum.6.010307}. Related developments have addressed feedforward readout mitigation and the characterization of MCM-specific noise~\cite{cj89-4h5t,HinesProctor2025,ZhangChenLiuJiang2025}.

Despite these advances, applying ZNE to mid-circuit measurements requires a means of systematically varying measurement-noise strengths while preserving the outcome-conditioned quantum operations relevant to subsequent feedback. Unlike unitary gates, nonunitary measurements do not naturally admit conventional gate-folding techniques for controlled noise amplification. It is therefore useful to investigate whether the noisy operations employed for instrument-level PEC can also be used to generate prescribed levels of measurement noise. However, the existence of a PEC reconstruction does not necessarily imply that the same operations can realize noise amplification through physical probabilistic sampling, since PEC permits signed quasiprobability coefficients whereas stochastic amplification requires nonnegative sampling probabilities. This motivates a framework that establishes the relationship between instrument-level reconstruction and controlled noise amplification and identifies the conditions under which both can be implemented using the same experimentally accessible operations.

In this work, we develop a quantum-instrument framework for PEC and controlled noise amplification of mid-circuit measurements. Our approach characterizes errors in classical measurement outcomes and post-measurement quantum states, including their correlations, while preserving the logical outcomes used for feedforward. We derive the reconstruction of ideal measurement operations using experimentally accessible noisy operations and establish the conditions for physically realizable stochastic noise amplification using the same operations.

Technically, the main contribution of our framework is to establish an affine relation between the coefficients used for PEC reconstruction and those used for measurement-noise scaling. Denoting the PEC coefficients by $\boldsymbol{\gamma}$ and the noise-scaling coefficients by $\boldsymbol{\alpha}(\lambda)$, we obtain
\begin{equation}
\label{eq:intro_meas_affine}
\boldsymbol{\alpha}(\lambda)
=
(1-\lambda)\boldsymbol{\gamma}
+
\lambda\mathbf e_0,
\end{equation}
where $\mathbf e_0$ selects the undressed noisy operation. The parameter $\lambda$ interpolates between the ideal reconstructed instrument at $\lambda=0$ and the native noisy instrument at $\lambda=1$, while $\lambda>1$ specifies a target with amplified stochastic measurement noise. For separately calibrated preparation noise, an analogous relation holds:
\begin{equation}
\label{eq:intro_prep_affine}
\boldsymbol{\beta}(\lambda)
=
(1-\lambda)\boldsymbol{\eta}
+
\lambda\mathbf e_0,
\end{equation}
where $\boldsymbol{\eta}$ and $\boldsymbol{\beta}(\lambda)$ denote the preparation PEC and noise-scaling coefficients, respectively. These relations establish a direct algebraic connection between error cancellation and noise amplification using the same dressed operations. However, a valid PEC reconstruction does not necessarily imply that noise amplification can be physically realized, as the latter additionally requires nonnegative sampling coefficients and a valid scaled error distribution. Our framework identifies the conditions under which such amplification is feasible within the specified dressed-operation protocol.

We validate the proposed framework numerically and through a proof-of-principle demonstration on a 20-qubit superconducting quantum processor developed at the Korea Research Institute of Standards and Science (KRISS). We first perform numerical simulations of iterative quantum phase estimation under various measurement-error models, showing that instrument-level PEC can mitigate post-measurement state disturbances that conventional readout correction cannot address, while ZNE can reduce measurement-induced errors through controlled noise amplification. We then implement a two-qubit adaptive phase-estimation circuit on the KRISS processor, incorporating mid-circuit measurements, feedforward, and qubit reuse. Using experimentally calibrated measurement errors, we implement PEC reconstruction and stochastic noise scaling for ZNE within the same dressed-operation framework. The experimental results show that both methods produce phase-expectation point estimates closer to the ideal reference than conventional readout correction, demonstrating the feasibility of the proposed approach on superconducting quantum hardware.

\section{Method}
\label{sec:method}
We formulate a Pauli-only framework for mitigating mid-circuit measurement errors in dynamic circuits in terms of outcome-resolved quantum instruments; state-preparation mitigation is treated as a complementary component for initialization and reuse. An instrument retains the classical measurement record and the corresponding conditional quantum operation, which are both necessary to specify the effect of a mid-circuit measurement on later feedforward. Under the stated twirling and computational-subspace assumptions, randomized compiling (RC) tailors the effective preparation and measurement noise to stochastic Pauli-$X$ models. The resulting error model admits a common set of experimentally accessible operations for probabilistic error cancellation (PEC) and zero-noise extrapolation (ZNE). Throughout this section, $a,b,c,k\in\{0,1\}$, and $\mathcal X^s(\rho)=X^s\rho X^s$ for $s\in\{0,1\}$, with $X^0=I$. In measurement operations, $b$ labels a pre-measurement bit flip, $a$ a post-measurement bit flip, and $k$ the logically corrected classical outcome used for feedforward.

\subsection{Effective Error Modeling via Randomized Compiling}
\label{sec:measurement_rc}
\emph{Origin and noise-tailoring principle.---}
RC was introduced to tailor the noise of quantum gates by inserting randomly chosen operations whose compensations preserve the intended computation~\cite{PhysRevA.94.052325}. In its simplest form, averaging a channel $\mathcal E$ over Pauli conjugations removes off-diagonal terms in its Pauli-transfer representation, yielding a stochastic Pauli channel. RC therefore does not eliminate physical noise; instead, it replaces a difficult-to-characterize effective error process with a structured one. For an arbitrary qubit channel, the Pauli twirl is
\begin{equation}
\mathcal E_{\rm PT}=\frac14\sum_{P\in\{I,X,Y,Z\}}\mathcal P\circ\mathcal E\circ\mathcal P, ~~\mathrm{where}
~~\mathcal P(\rho)=P\rho P^{\dagger}.
\end{equation}
The extension to quantum measurements is more subtle because a measurement specifies both a reported classical outcome and a conditional output state. Randomized compiling for subsystem measurements was developed in Ref.~\cite{beale2023randomized}; measurement RC was subsequently used for quasiprobabilistic readout correction in adaptive circuits~\cite{PRXQuantum.6.010307}. Here we use the resulting instrument-level noise representation to treat disturbances to the measured quantum state along with errors affecting feedforward.

\emph{Instrument and randomized protocol.---}
For a finite outcome set $\Omega$, a quantum instrument is a family
$\mathfrak I=\{\mathcal I_k\}_{k\in\Omega}$ of completely positive,
trace-nonincreasing linear maps from input to output operators such that
$\sum_{k\in\Omega}\mathcal I_k$ is trace-preserving. The operational
probability and normalized conditional output are, respectively,
\begin{equation}
\label{eq:instrument_definition}
  p(k )=\operatorname{tr}[\mathcal I_k(\rho)]~~\mathrm{and}~~
  \sigma_k= \frac{ 1}{p(k)} \mathcal I_k(\rho),~~\mathrm{where}~~
   p(k)>0.
\end{equation}
The corresponding classical--quantum channel is
$\widehat{\mathcal I}(\rho)=\sum_k\mathcal I_k(\rho)\otimes\kb{k}_C$. We call each outcome-conditioned operation $\mathcal I_k$ a \emph{measurement branch}; the term \emph{branchwise} means that an identity holds separately for every outcome $k$. The instrument is the outcome-indexed family, whereas its flagged
channel is one convenient representation that retains the outcome record.
A Kraus representation
$\mathcal I_k(\rho)=\sum_\alpha K_{k\alpha}\rho K_{k\alpha}^\dagger$
is not unique.

\begin{figure*}[t]
\includegraphics[width=0.98\textwidth]{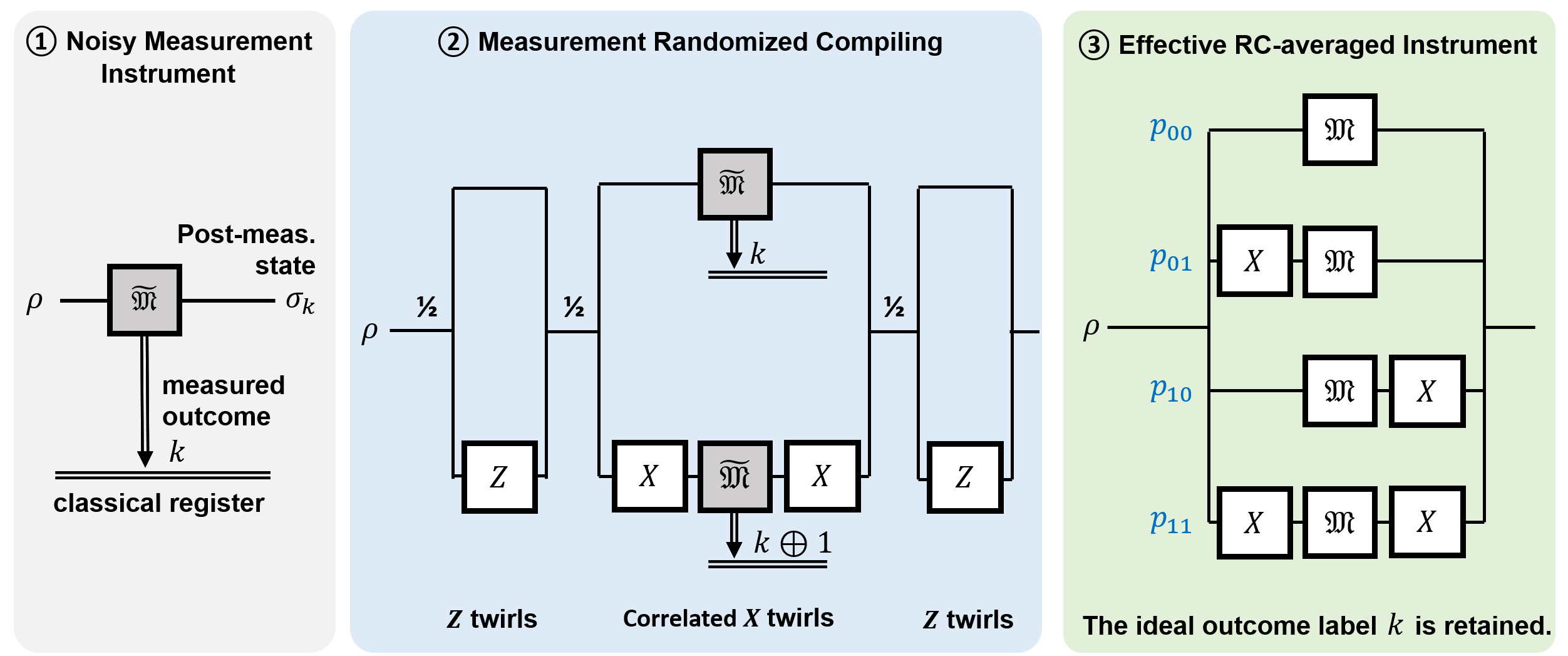}
\caption{Instrument-level randomized compiling for a binary qubit measurement is shown. Independent input and output $Z$ twirls suppress computational-basis coherences, while a correlated $X$ twirl symmetrizes the two measurement branches. Recording the outcome as $k \oplus 1$ when the $X$ twirl is applied yields an averaged instrument with stochastic weights $p_{ab}$ for effective post-measurement ($a$) and pre-measurement ($b$) flips. These labels need not identify unique microscopic fault times.}
\label{fig:RC}
\end{figure*}

We first consider sequential single-qubit measurements followed by reinitialization and reuse, and assume that measurement noise is independent of unmeasured qubits. Extensions involving spectator degrees of freedom require an additional jointly covariant error model and are outside the single-qubit derivation presented here. Let $\mathfrak M=\{\mathcal M_k\}_{k=0,1}$ denote the ideal projective instrument and $\widetilde{\mathfrak M}=\{\widetilde{\mathcal M}_k\}_{k=0,1}$ the physical noisy instrument. Their outcome-conditioned operations and the flagged ideal channel are
\begin{equation}
\label{eq:ideal_instrument}
\begin{aligned}
\mathcal M_k(\rho)&=\Pi_k\rho\Pi_k,~~\mathrm{where}
~~ \Pi_k=\ket{k}\!\bra{k}\\
\mathrm{and}~~\widehat{\mathcal M}(\rho)&=\sum_{k=0}^1\mathcal M_k(\rho)\otimes\kb{k}_C.
\end{aligned}
\end{equation}
We write $\mathcal Z^s(\rho)=Z^s\rho Z^s$, analogously to $\mathcal X^s$, with $s\in\{0,1\}$. The measurement RC protocol (Fig.~\ref{fig:RC}) samples three independent, uniformly distributed bits $z_1,z_2,x$. The first two prescribe $Z^{z_1}$ before and $Z^{z_2}$ after the physical measurement. The third prescribes $X^x$ before and after the measurement, accompanied by a classical correction $k=k_{\rm phys}\oplus x$. The RC-averaged instrument $\overline{\mathfrak M}=\{\overline{\mathcal M}_k\}_{k=0,1}$ has outcome-conditioned maps
\begin{equation}
\label{eq:explicit_measurement_twirl}
\overline{\mathcal M}_k=\frac18\sum_{z_1,z_2,x\in\{0,1\}}
\mathcal Z^{z_2}\circ\mathcal X^x\circ
\widetilde{\mathcal M}_{k\oplus x}\circ
\mathcal X^x\circ\mathcal Z^{z_1}.
\end{equation}
Equation~\eqref{eq:explicit_measurement_twirl} specifies the average instrument over random configurations; individual experimental shots use a definite sampled configuration. For the ideal instrument, $\mathcal X^x\circ\mathcal M_{k\oplus x}\circ\mathcal X^x=\mathcal M_k$; moreover, $\mathcal Z$ leaves its input and output projectors invariant. Thus the random operations preserve the ideal measurement while tailoring the noise. In particular, the physical bit $k_{\rm phys}$ must not be passed unchanged to feedforward when $x=1$: the corrected logical bit is $k$.

\emph{Reduction to a four-parameter error model.---}
The independent pre- and post-measurement $Z$ averages eliminate coherences between computational-basis operators in the effective instrument. Correlated $X$ twirling then symmetrizes the two outcome-conditioned operations. This argument assumes a binary-outcome, qubit-to-qubit completely positive instrument and does not include leakage outside the computational subspace. For a single qubit, the resulting conditional map has four nonnegative weights $p_{ab}$, with $a$ and $b$ indexing post- and pre-measurement $X$ errors, respectively. The first main result makes this reduction precise, including its domain of validity.

\begin{proposition}[Single-qubit instrument tailoring]\label{prop:tailoring}
Let $\{\widetilde{\mathcal M}_k\}_{k=0,1}$ be a binary-outcome, completely positive, qubit-to-qubit instrument whose sum is trace preserving. Assume that the specified independent pre- and post-measurement $Z$ twirls and the correlated $X$ twirl are applied and averaged, including the logical outcome correction $k=k_{\rm phys}\oplus x$. On the computational subspace, the averaged instrument has the form in Eq.~\eqref{eq:RC_channel}, with $p_{ab}\geq0$ and $\sum_{ab}p_{ab}=1$.
\end{proposition}
\noindent\emph{Proof.} The two independent $Z$ averages project the input and output operator spaces onto their diagonal sectors, giving the diagonal conditional map $\mathcal D_k(\rho)=\sum_{i,j=0}^1 t_{jki}\langle i|\rho|i\rangle\ket{j}\!\bra{j}$ with $t_{jki}\geq0$ by complete positivity. Averaging the correlated $X$ action enforces $t_{jki}^{\mathrm{RC}}=t_{j\oplus1,k\oplus1,i\oplus1}^{\mathrm{RC}}$, so the entries depend only on $a=j\oplus k$ and $b=i\oplus k$, i.e., $t_{jki}^{\mathrm{RC}}=p_{j\oplus k,i\oplus k}$. For fixed input $i$, the transformation $(j,k)\leftrightarrow(a,b)$ is bijective, so trace preservation yields $\sum_{k,j}t_{jki}^{\mathrm{RC}}=\sum_{a,b}p_{ab}=1$. Substitution into the diagonal conditional map gives Eq.~\eqref{eq:RC_channel} for every $k$. $\square$

Explicitly,
\begin{equation}
\label{eq:RC_channel}
\overline{\mathcal M}_k=\sum_{a,b\in\{0,1\}}p_{ab}\,
\mathcal X^a\circ\mathcal M_k\circ\mathcal X^b,
~~\mathrm{where}~~ \sum_{a,b}p_{ab}=1.
\end{equation}
Equivalently, the outcome-conditioned maps admit the Kraus representation
\begin{equation}
\begin{aligned}
E_{abk}&=\sqrt{p_{ab}}\ket{k\oplus a}\!\bra{k\oplus b},\\
\overline{\mathcal M}_k(\rho)
&=\sum_{a,b\in\{0,1\}}E_{abk}\rho E_{abk}^{\dagger}\\
\mathrm{and}~~\widehat{\overline{\mathcal M}}(\rho)
&=\sum_{k=0}^{1}\overline{\mathcal M}_k(\rho)\otimes\kb{k}_C.
\end{aligned}
\end{equation}
The Kraus operators obey the completeness relation
$\sum_{k,a,b}E_{abk}^{\dagger}E_{abk}
=\sum_{a,b}p_{ab}I=I$, so their outcome-conditioned maps form a normalized instrument.
In terms of the input density matrix, this representation reads
\begin{equation}
\overline{\mathcal M}_k(\rho)=\sum_{a,b}p_{ab}\,
\bra{k\oplus b}\rho\ket{k\oplus b}\,
\kb{k\oplus a}.
\end{equation}
For any input $\rho$, $\mathrm{tr}[\overline{\mathcal M}_k(\rho)]$ gives the distribution of the corrected logical record, while Eq.~\eqref{eq:instrument_definition} applied to $\overline{\mathcal M}_k$ gives the conditional output state. The index $b$ identifies which input-basis population contributes to logical outcome $k$; $a$ identifies the final quantum output relative to the ideal post-measurement state $\ket{k}$. Hence $p_{01}$ captures a readout-associated pre-measurement flip, $p_{10}$ an effective post-measurement $X$ flip, and $p_{11}$ both effects. Crucially, the classical register in Eq.~\eqref{eq:RC_channel} is labeled by the corrected outcome, not an uncorrected physical detector bit. Note that this four-parameter form applies to the RC-averaged effective instrument rather than a generic unaveraged measurement.

For later error scaling, define the scalar total nonidentity error probability $p_{\mathrm{err}}:=1-p_{00}=p_{01}+p_{10}+p_{11}$. We abbreviate this scalar as $p:=p_{\mathrm{err}}$ in the single-qubit formulas below. When $p>0$, Eq.~\eqref{eq:RC_channel} may equivalently be separated into ideal and erroneous contributions:
\begin{equation}
\begin{aligned}
\overline{\mathcal M}_k&=(1-p)\mathcal M_k+p\mathcal N_k,
~~\mathrm{where}~~\mathcal N_k=\frac1p\sum_{(a,b)\ne(0,0)}p_{ab}\,
\mathcal X^a\circ\mathcal M_k\circ\mathcal X^b.
\end{aligned}
\end{equation}
The four-parameter expression remains well defined when $p=0$. The characterization procedure for $p_{ab}$ is described in Appendix~\ref{app:characterize}.

\begin{figure*}[t]
\includegraphics[width=0.98\textwidth]{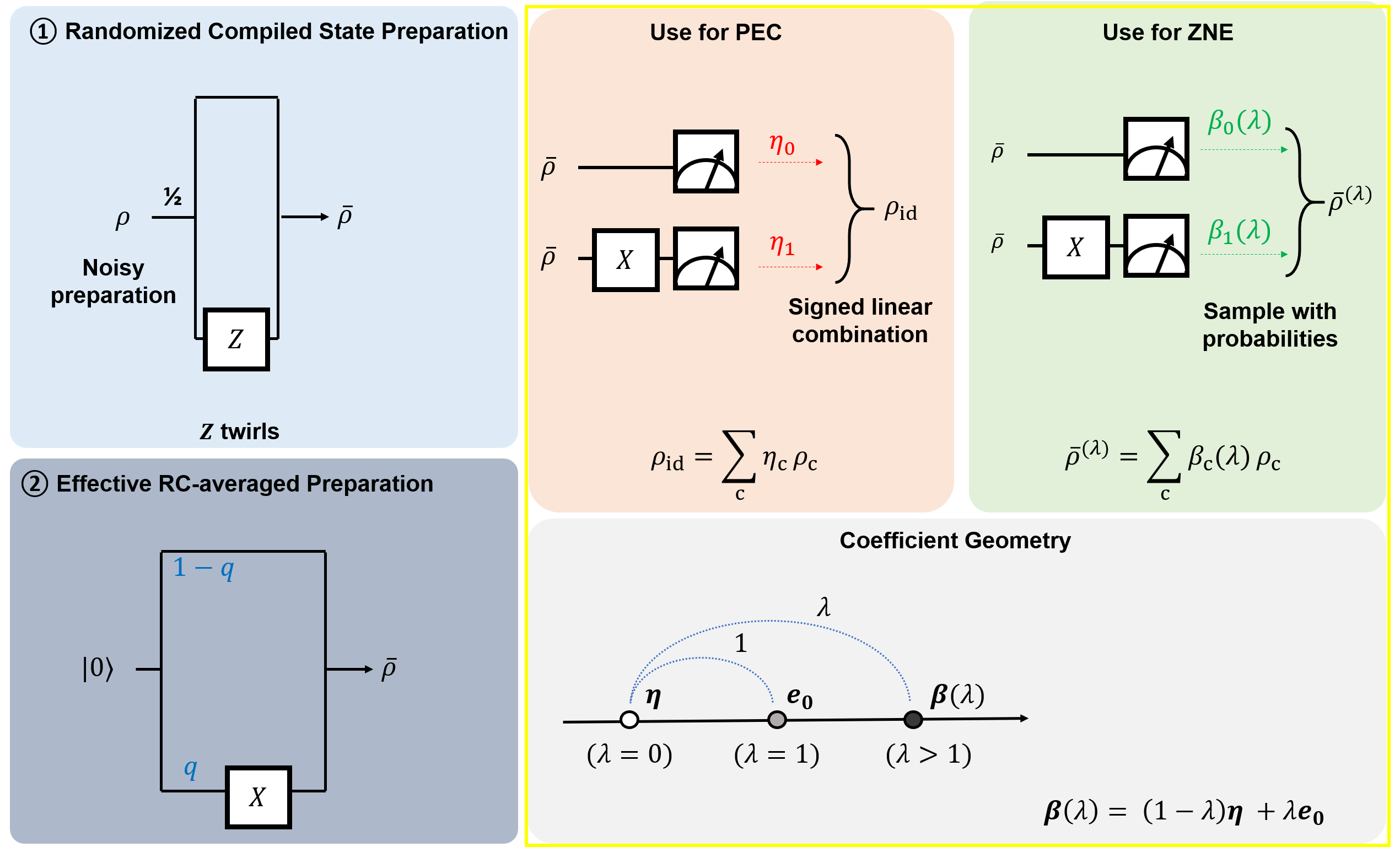}
\caption{Preparation-error tailoring and mitigation is shown. A random $Z$ twirl maps imperfect $|0\rangle$ preparation to $(1-q)|0\rangle\langle0|+q|1\rangle\langle1|$. PEC reconstructs the ideal state preparation using signed coefficients $\eta_c$ over $X$-dressed noisy preparations. ZNE samples the same operations to realize prescribed amplified stochastic errors when every $\beta_c(\lambda)\geq0$. The coefficients obey $\boldsymbol\beta(\lambda)=(1-\lambda)\boldsymbol\eta+\lambda\mathbf e_0$: $\lambda=0$ gives signed reconstruction, $\lambda=1$ selects the native preparation, and $\lambda>1$ specifies amplification.}
\label{fig:PRC}
\end{figure*}

\emph{Preparation twirling.---}
An analogous random $Z$ operation removes off-diagonal components of the initial state $\rho$. For the target preparation $\rho_{\rm id}=\kb{0}$, one obtains, see Fig. \ref{fig:PRC}
\begin{equation}
\label{eq:PRC_state}
\overline\rho=\frac12(\rho+Z\rho Z)
=(1-q)\rho_{\rm id}+q\mathcal X(\rho_{\rm id}),~~\mathrm{where}~~ q=\bra{1}\rho\ket{1}.
\end{equation}
Thus both preparation and the RC-averaged measurement can be described by stochastic Pauli-$X$ operations. This is the representation exploited below by PEC and ZNE; the preparation parameter $q$ is characterized separately.

\subsection{Separating Readout Errors from Post-Measurement State Disturbance}
\label{sec:readout_limit}
Measurement backaction broadly denotes the change in a quantum state associated with a measurement, including the projection intrinsic to an ideal measurement. Here, we use \emph{post-measurement state disturbance} for additional errors relative to the ideal conditional state. In the RC-averaged stochastic model, the label $a=1$ represents an effective Pauli-$X$ error after the ideal projection; it does not characterize all physical mechanisms of measurement backaction. In particular, $p_{10}$ isolates a post-measurement flip without a pre-measurement flip, while $p_{10}+p_{11}$ is the marginal probability of a post-measurement flip.\par
A useful diagnostic is a noise model in which every logical outcome is correct, but the measured qubit is disturbed after projection. Set $p_{00}=1-r$, $p_{10}=r$, and $p_{01}=p_{11}=0$, with $0<r<1/2$. Then
\begin{equation}
\label{eq:pure_post}
\overline{\mathcal M}_k(\rho)=(1-r)\Pi_k\rho\Pi_k
+rX\Pi_k\rho\Pi_kX.
\end{equation}
The outcome probability agrees exactly with the ideal instrument,
$\mathrm{tr}[\overline{\mathcal M}_k(\rho)]=\mathrm{tr}[\Pi_k\rho]$,
so that the readout confusion matrix is the identity. Conditional on outcome $k$, we have a resulting state $(1-r)\kb{k}+r\kb{k\oplus1}$. A following gate, reset, or reuse operation acting on that qubit can therefore produce a biased downstream observable. In this case, correction based only on the outcome-assignment matrix makes no change, although there remains an error in the instrument.

To illustrate instrument-level cancellation, define the undressed noisy instrument $\mathfrak B_{00}=\overline{\mathfrak M}=\{\overline{\mathcal M}_k\}_{k=0,1}$ and its post-measurement Pauli-dressed counterpart $\mathfrak B_{10}=\{\mathcal X\circ\overline{\mathcal M}_k\}_{k=0,1}$, with $\mathcal X(\rho)=X\rho X$. Both are families of outcome-conditioned maps, with the same corrected logical outcome label $k$. Then, for the $p_{10}$-only model, the identity
\begin{equation}
\label{eq:pure_post_pec}
\mathfrak M=\frac{1-r}{1-2r}\,\mathfrak B_{00}
-\frac{r}{1-2r}\,\mathfrak B_{10}
\end{equation}
holds separately for each outcome and reconstructs the ideal outcome-conditioned maps as a signed estimator identity, enabling unbiased downstream expectation values within the specified model. It does not physically restore an ideal post-measurement state.

\noindent\textit{Illustrative measurement--reset diagnostic.} Consider the $p_{10}$-only model in Eq.~\eqref{eq:pure_post}, in which the logical measurement outcome $k$ is correct but the conditional quantum state undergoes a post-measurement $X$ flip with probability $r$. Following outcome $k$, apply the conditional reset $X^k$, which ideally prepares the measured qubit in $\ket{0}$. With the state error, the output after reset is instead given as, $\rho_{\rm reset}=(1-r)\kb{0}+r\kb{1}$, which is independent of $k$. Assuming an ideal final $Z$ measurement and ideal reset gate, the expectation value is $E_{\rm raw}(r)=\mathrm{tr}[Z\rho_{\rm reset}]=1-2r$, whereas $E_{\rm id}=1$. Because the first measurement's classical record is error-free, assignment-matrix inversion for that measurement cannot remove this residual state error. In contrast, the outcome-wise PEC identity in Eq.~\eqref{eq:pure_post_pec} gives $E_{\rm PEC}=1$ at the estimator level.

For ZNE, scale the post-measurement flip probability to $\lambda r$ within the same stochastic model. The reset expectation then becomes $E(\lambda)=1-2\lambda r$, so first-order Richardson extrapolation at scales $\{1,2\}$ gives $2E(1)-E(2)=1$ exactly. This scaling is experimentally sampleable from the two dressed instruments only when the coefficients $\alpha_{00}(\lambda)=[1-(1+\lambda)r]/(1-2r)$ and $\alpha_{10}(\lambda)=(\lambda-1)r/(1-2r)$ are nonnegative. In the assumed regime $0\le r<1/2$, scale $\lambda=2$ is admissible for $0\le r\le 1/3$. The exact extrapolation here follows from the linear dependence on $\lambda$ and need not hold for a general adaptive circuit.

The $p_{10}$-only diagnostic isolates information that is absent from an assignment matrix: the corrected logical record can have ideal statistics even though the conditional state passed to reset or reuse is erroneous. Assignment-only correction is therefore the identity in this example, while branchwise instrument cancellation recovers the downstream ideal expectation at the estimator level. This comparison distinguishes complete instrument mitigation from readout-only correction.

\subsection{Probabilistic Error Cancellation}\label{sec:pec}
PEC reconstructs each ideal outcome-conditioned measurement operation as a signed linear combination of experimentally accessible noisy operations. This is \emph{error mitigation}, not physical quantum error correction: a sampled circuit still undergoes measurement errors, and PEC does not restore the ideal conditional state in an individual run. Instead, appropriately weighted averages over different noisy circuit configurations reproduce ideal expectation values within the calibrated stochastic model. The identity must hold separately for every corrected logical outcome $k$, since later gates and resets may depend on that outcome. This outcome-wise reconstruction is stronger than correcting marginal outcome probabilities alone: it also accounts for the unnormalized conditional quantum operation passed to subsequent circuit steps.

Generalizing the two dressed instruments introduced in Sec.~\ref{sec:readout_limit}, for each fixed corrected logical outcome $k$ define the four Pauli-dressed basis maps
\begin{equation}
\label{eq:basis_channel}
\mathcal B_{ab,k}=\mathcal X^a\circ\overline{\mathcal M}_k\circ\mathcal X^b.
\end{equation}
The subscript $k$ makes the outcome-conditioned map explicit. In an implementation, the logical register label must also be tracked through the randomized-compiling and Pauli-dressing corrections; the displayed composition specifies the quantum map conditioned on that logical label. The label $b$ specifies an additional $X$ before the noisy measurement and $a$ an additional $X$ after it. Thus $(a,b)=(0,0)$ leaves the calibrated instrument unchanged, while the other three choices change its Pauli frame. These extra operations are known dressings of the \emph{same noisy instrument}, not four independently characterized measurement devices. The baseline map is $\mathcal B_{00,k}=\overline{\mathcal M}_k$. Order the error labels as $(00,01,10,11)$ and collect the corresponding maps into column vectors:
\begin{equation}
\label{eq:basis_expansion}
\mathbf B_k=P\mathbf M_k^X,~~\mathrm{where}~~
P=\begin{pmatrix}
p_{00}&p_{01}&p_{10}&p_{11}\\
p_{01}&p_{00}&p_{11}&p_{10}\\
p_{10}&p_{11}&p_{00}&p_{01}\\
p_{11}&p_{10}&p_{01}&p_{00}
\end{pmatrix}.
\end{equation}
Here $\mathbf B_k=(\mathcal B_{ab,k})_{ab}^T$ and $\mathbf M_k^X=(\mathcal X^a\circ\mathcal M_k\circ\mathcal X^b)_{ab}^T$, both in the stated order. More explicitly,
$\mathcal B_{ab,k}=\sum_{c,d}p_{cd}
(\mathcal X^{a\oplus c}\circ\mathcal M_k\circ\mathcal X^{b\oplus d})$,
so that the coefficient of the ideal basis map with label $(e,f)$ is
$p_{a\oplus e,\,b\oplus f}$. This establishes Eq.~\eqref{eq:basis_expansion} directly. Each row of $P$ is consequently a permutation of the same four calibrated probabilities: applying a dressing changes the error label by bitwise addition modulo two, but does not change the underlying error distribution. This group-convolution structure is why a single matrix relates all dressed maps to their ideal counterparts. Operationally, the classical register is the corrected logical result $k=k_{\rm phys}\oplus x$ of the inner RC measurement, and external Pauli dressing does not itself relabel this register. Conditional feedforward must use that logical result, together with any explicitly prescribed feedforward-frame updates, rather than the raw detector bit.

Let $\mathbf e_0=(1,0,0,0)^T$ and $\boldsymbol\gamma=(\gamma_{00},\gamma_{01},\gamma_{10},\gamma_{11})^T$. To extract the undressed ideal map, consider a linear combination $\sum_{a,b}\gamma_{ab}\mathcal B_{ab,k}$. In the ordered ideal basis $\mathbf M_k^X$, its coefficient vector is $P^T\boldsymbol\gamma=P\boldsymbol\gamma$, since $P$ is symmetric. Requiring coefficient one for $\mathcal M_k$ and zero for the other three ideal basis maps gives $P\boldsymbol\gamma=\mathbf e_0$. Provided $P$ is invertible, the ideal outcome-conditioned map is reconstructed as
\begin{equation}
\mathcal M_k=\sum_{a,b}\gamma_{ab}\mathcal B_{ab,k},~~\mathrm{where}~~
\boldsymbol\gamma=P^{-1}\mathbf e_0.
\end{equation}
The Walsh--Hadamard transform diagonalizes $P$. Define the four Fourier
coefficients of the error distribution by
\begin{equation}
\label{eq:single_fourier}
 d_{uv}=\sum_{a,b\in\{0,1\}}(-1)^{ua+vb}p_{ab},~~\mathrm{where}~~ u,v\in\{0,1\}.
\end{equation}
The numbers $d_{uv}$ quantify how strongly each binary parity mode of the error distribution survives the noise. In particular, $d_{00}=1$ because the probabilities sum to one. Let $H_4$ be the $4\times4$ Walsh--Hadamard matrix with rows indexed by $(u,v)$, columns indexed by $(a,b)$, and entries $(H_4)_{(u,v),(a,b)}=(-1)^{ua+vb}$. Then, one has
$H_4H_4^T=4I$ and
$P=\frac14 H_4^T\operatorname{diag}(d_{00},d_{01},d_{10},d_{11})H_4$, from which it holds that $P^{-1}=\frac14 H_4^T\operatorname{diag}(d_{uv}^{-1})H_4$ whenever all $d_{uv}\neq0$, and the PEC weights are
\begin{equation}
\label{eq:gamma_coefficients}
 \gamma_{ab}=\frac14\sum_{u,v\in\{0,1\}}
 \frac{(-1)^{ua+vb}}{d_{uv}}.
\end{equation}

\paragraph*{Finite-group notation.}
Let $G$ be a finite abelian group, written additively with identity $0$, and let $k$ label an outcome branch. Let $p:G\to[0,1]$ be a normalized error distribution, $\mathcal A_{g,k}$ the ideal operation with error-label dressing $g$ (so $\mathcal A_{0,k}$ is the desired ideal branch), and $\mathcal B_{g,k}$ the experimentally accessible noisy dressed operation. We assume the group-covariant relation
\begin{equation}\label{eq:group_covariance}
\mathcal B_{g,k}=\sum_{h\in G}p(h)\mathcal A_{g+h,k}, \quad g\in G.
\end{equation}
Define the convolution matrix $C$ by $C_{t,g}=p(t-g)$, with row index $t$ and column index $g$, and let $\mathbf e_0\in\mathbb R^{|G|}$ denote the unit vector supported on the group identity. A character $\chi:G\to\mathbb C$ is a homomorphism into the complex unit circle; we adopt the Fourier convention $\widehat p(\chi)=\sum_{h\in G}p(h)\overline{\chi(h)}$.
\begin{theorem}[Finite-group convolution reconstruction]\label{thm:convolution}
If $\widehat p(\chi)\neq0$ for every character $\chi$ of $G$, then $C$ is invertible and the unique coefficient vector in the group-convolution representation that reconstructs every ideal branch is $\boldsymbol\gamma=C^{-1}\mathbf e_0$, namely
\begin{equation}\label{eq:group_pec}
\mathcal A_{0,k}=\sum_{g\in G}\gamma_g\mathcal B_{g,k}\quad\text{for every }k.
\end{equation}
For the present binary Pauli model, $G=\mathbb Z_2^2$ and $C=P=P^T$.
\end{theorem}
\noindent\emph{Proof.} Expanding $\sum_g\gamma_g\mathcal B_{g,k}$ with Eq.~\eqref{eq:group_covariance}, the coefficient multiplying $\mathcal A_{t,k}$ equals $\sum_g p(t-g)\gamma_g=(C\boldsymbol\gamma)_t$. For any character $\chi$, one has $(C\chi)(t)=\chi(t)\sum_h p(h)\overline{\chi(h)}=\widehat p(\chi)\chi(t)$. The characters form a basis, so $C$ is invertible precisely when every $\widehat p(\chi)$ is nonzero. Hence $C\boldsymbol\gamma=\mathbf e_0$ has the unique solution $\boldsymbol\gamma=C^{-1}\mathbf e_0$, proving the branchwise identity. $\square$

The uniqueness statement concerns coefficient vectors in the formal group-convolution representation. Since distinct ideal branches may be linearly dependent as superoperators, it does not imply uniqueness of a physical linear decomposition. The theorem assumes the group-covariant dressed family; it does not establish that arbitrary instruments admit such tailoring. For the present binary model, the nonzero-Fourier condition is necessary and sufficient for invertibility of the full four-dimensional convolution matrix, though less demanding observable-specific reconstructions may exist.

The convolution matrix is invertible precisely when all four $d_{uv}$ are nonzero; this is a sufficient distribution-level condition for the displayed reconstruction. A vanishing Fourier coefficient means that the corresponding parity information is completely lost in this distribution-level representation, so the displayed inverse does not exist. Small but nonzero $|d_{uv}|$ make inversion ill-conditioned: the coefficients $\gamma_{ab}$ can become large in magnitude even though the reconstruction is algebraically exact. This increases the statistical cost and makes the result sensitive to calibration uncertainty. This notation extends directly to multiqubit error
when joint group covariance holds.
\begin{figure*}[t]
\includegraphics[width=0.98\textwidth]{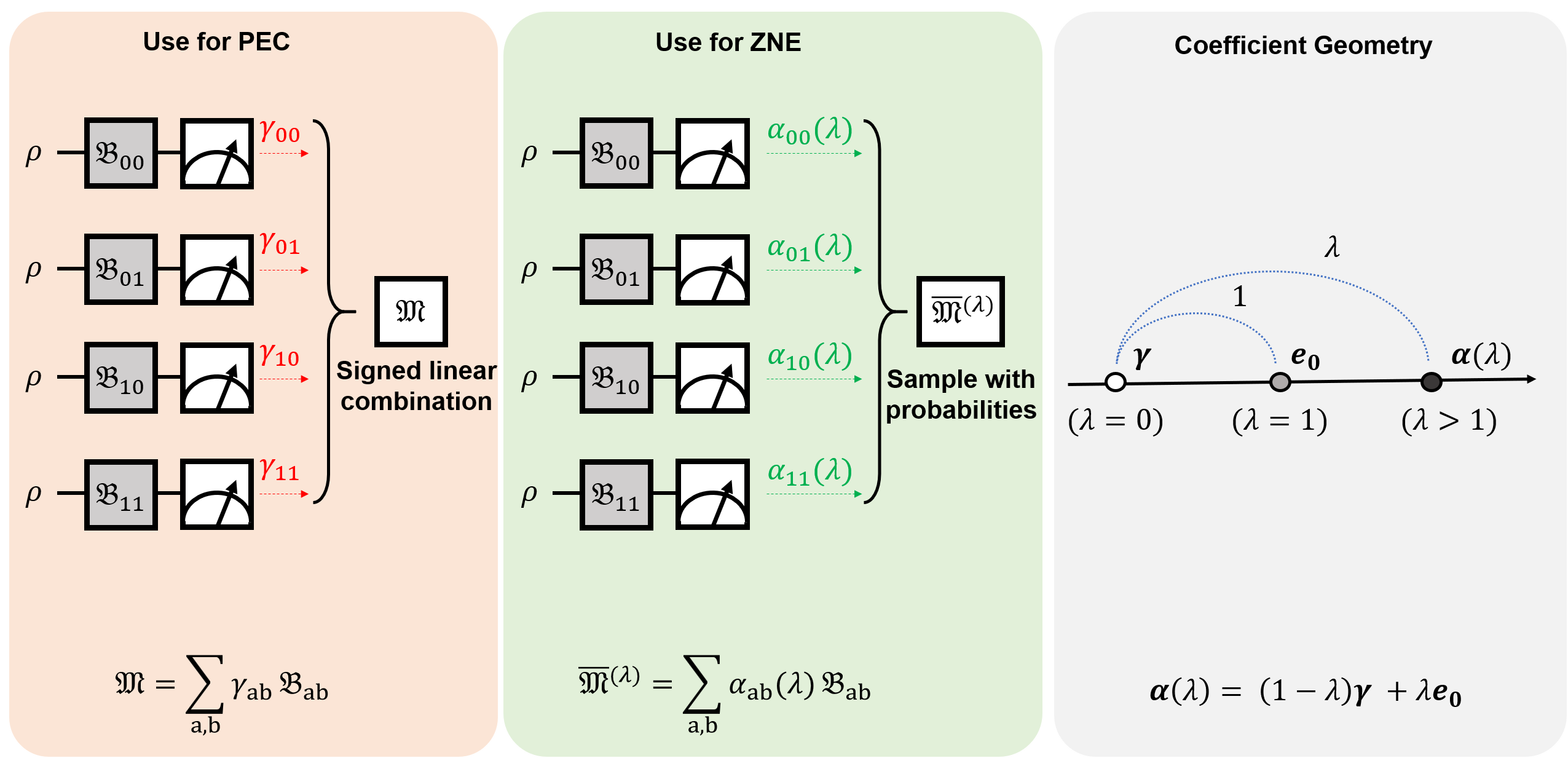}
\caption{The dressed-operation representation for PEC and ZNE is summarized. PEC reconstructs each ideal instrument branch using signed coefficients $\gamma_{ab}$ over Pauli-dressed noisy operations. ZNE samples the same operations to realize prescribed amplified stochastic errors when every $\alpha_{ab}(\lambda)\geq0$. The coefficients obey $\boldsymbol\alpha(\lambda)=(1-\lambda)\boldsymbol\gamma+\lambda\mathbf e_0$: $\lambda=0$ gives signed reconstruction, $\lambda=1$ selects the native instrument, and $\lambda>1$ specifies amplification.}
\label{fig:PEC1}
\end{figure*}

The same coefficients apply to every outcome $k$. In instrument notation,
\begin{equation}
\label{eq:pec_instrument}
\mathfrak M=\sum_{a,b}\gamma_{ab}\,\mathfrak B_{ab},~~\mathrm{where}~~
\mathfrak B_{ab}=\{\mathcal B_{ab,k}\}_{k=0,1},
\end{equation}
where the sum of families means an equality for each $k$. Equivalently,
$\widehat{\mathcal M}(\rho)=\sum_{k,a,b}\gamma_{ab}\mathcal B_{ab,k}(\rho)\otimes\kb{k}_C$
(Fig.~\ref{fig:PEC1}). The signed combination in Eq.~\eqref{eq:pec_instrument}
is an estimator-level quasiprobability identity, not in general a physical
instrument sampled with nonnegative probabilities.

Preparation PEC uses the basis states $\rho_c=\mathcal X^c(\overline\rho)$, giving
\begin{equation}
\rho_{\mathrm{id}}=\sum_{c=0}^1\eta_c\rho_c,~~\mathrm{where}~~
\label{eq:eta_coefficients}
\boldsymbol\eta=\begin{pmatrix}\eta_0\\\eta_1\end{pmatrix}
=\frac{1}{1-2q}\begin{pmatrix}1-q\\-q\end{pmatrix}.
\end{equation}
Indeed, $\overline\rho=(1-q)\rho_{\mathrm{id}}+q\mathcal X(\rho_{\mathrm{id}})$ and $\mathcal X(\overline\rho)=q\rho_{\mathrm{id}}+(1-q)\mathcal X(\rho_{\mathrm{id}})$. Solving these two linear equations for $\rho_{\mathrm{id}}$ gives the coefficients above. This inverse exists for $q\neq1/2$; in practice the low-error regime $q<1/2$ is considered (Fig.~\ref{fig:PRC}). The same inverse coefficients also determine exact preparation-noise scaling through Eq.~\eqref{eq:beta_affine}.

For a circuit with $N_{\rm P}$ preparations and $N_{\rm M}$ measurements, assume independently and identically distributed error instances. Define the quasiprobability norms and corresponding sampling distributions as
\begin{equation}
\Gamma_{\mathrm M}=\sum_{a,b}|\gamma_{ab}|,\quad
\pi_{ab}=\frac{|\gamma_{ab}|}{\Gamma_{\mathrm M}},\qquad
\Gamma_{\mathrm P}=\sum_c|\eta_c|~~\mathrm{and}~~
\pi_c=\frac{|\eta_c|}{\Gamma_{\mathrm P}}.
\end{equation}

{
\vspace{0.7em}
\noindent\textbf{Algorithm: PEC for $N_{\rm P}$ Preparations and $N_{\rm M}$ Measurements}
\vspace{0.3em}\hrule\vspace{0.3em}
\textbf{Input:} Target circuit, characterized $p_{ab},q$, and PEC weights $\boldsymbol\gamma,\boldsymbol\eta$ with norms $\Gamma_{\rm M},\Gamma_{\rm P}$.
\begin{enumerate}
\item Compute the sampling distributions $\pi_{ab}=|\gamma_{ab}|/\Gamma_{\rm M}$ and $\pi_c=|\eta_c|/\Gamma_{\rm P}$.
\item For each trajectory, initialize the sign factor $S=1$.
\item For every preparation, implement RC (random $Z$), sample $c$ with probability $\pi_c$, insert $X^c$, and update $S \leftarrow S\operatorname{sgn}(\eta_c)$.
\item For each MCM, implement measurement RC, sample $(a,b)$ with probability $\pi_{ab}$, insert $X^b$ before and $X^a$ after the RC measurement, track the corrected logical outcome for feedforward, and update $S \leftarrow S\operatorname{sgn}(\gamma_{ab})$.
\item Execute the resulting circuit to measure the observable $O$, and average the weighted outcome $\widehat O_{\rm PEC}=S\,\Gamma_{\rm P}^{N_{\rm P}}\Gamma_{\rm M}^{N_{\rm M}}O$ over trajectories.
\end{enumerate}
\vspace{0.3em}\hrule\vspace{0.5em}
}

\noindent\textbf{PEC sampling prescription.} For each trajectory, initialize $S=1$. Independently sample each preparation dressing $c$ with probability $\pi_c$ and each MCM dressing $(a,b)$ with probability $\pi_{ab}$; implement their Pauli frames with the logical MCM outcome tracked through feedforward. Multiply $S$ by the signs of the selected $\eta_c$ and $\gamma_{ab}$. For a measured observable $O$, average the weighted outcome
\begin{equation}
\widehat O_{\rm PEC}=S\,\Gamma_{\rm P}^{N_{\rm P}}\Gamma_{\rm M}^{N_{\rm M}} O.
\end{equation}
\noindent To see why this procedure is unbiased, note that a dressing drawn with probability $|\gamma_{ab}|/\Gamma_{\mathrm M}$ and multiplied by $\Gamma_{\mathrm M}\operatorname{sgn}(\gamma_{ab})$ has an average contribution $\gamma_{ab}$ times the measured circuit response. The same argument applies to preparation with $\eta_c$. Averaging the signed and normalized trajectories therefore reproduces the linear PEC decomposition at every mitigated block. The weights apply to the \emph{entire} adaptive circuit: a sampled dressing changes a physical circuit trajectory, while the corrected logical measurement record still controls the intended feedforward. Repeated trajectories provide an unbiased estimator only within the calibrated, history-independent stochastic model, and must not be interpreted as physically implementing negative-probability instruments.

This estimator cancels errors exactly within the characterized model, but its sampling cost increases with the quasiprobability norms. The trajectory weight has magnitude $\Gamma_{\mathrm P}^{N_{\rm P}}\Gamma_{\mathrm M}^{N_{\rm M}}$, and its squared magnitude determines the corresponding worst-case variance bound, proportional to $\Gamma_{\mathrm P}^{2N_{\rm P}}\Gamma_{\mathrm M}^{2N_{\rm M}}$~\cite{RevModPhys.95.045005}.

\subsection{Zero-Noise Extrapolation}
\label{sec:zne}

ZNE estimates the ideal expectation value by evaluating the same circuit at several known noise strengths and extrapolating to zero noise. Unlike PEC, which directly reconstructs the ideal instrument using signed coefficients, ZNE requires \emph{physical} circuits with amplified noise. The expectation values measured at those noise strengths are subsequently combined using signed Richardson coefficients. We show below how the Pauli-dressed instruments used for PEC can also generate these amplified circuits, and under which conditions this can be done with ordinary probabilities.

\paragraph*{Target noise scaling.}
Recall that the scalar $p=p_{\mathrm{err}}=1-p_{00}=p_{01}+p_{10}+p_{11}$ denotes the total nonidentity error probability of the calibrated MCM instrument. We specify the desired noise strength by multiplying each nonidentity probability by the same factor $\lambda\geq1$, while adjusting the identity component to retain normalization:
\begin{equation}
\overline{\mathcal M}_k^{(\lambda)}
=(1-\lambda p)\mathcal M_k+\lambda p\mathcal N_k
=\sum_{a,b}p_{ab}^{(\lambda)}\,\mathcal X^a\circ\mathcal M_k\circ\mathcal X^b,
\end{equation}
where $p_{00}^{(\lambda)}=1-\lambda p$ and $p_{ab}^{(\lambda)}=\lambda p_{ab}$ for $(a,b)\ne(0,0)$. This construction retains the corrected logical outcome $k$ and specifies the entire outcome-conditioned operation, rather than scaling only the observed readout error. It is a \emph{model-defined} scaling of stochastic error probabilities, not a claim that gate durations or all physical error mechanisms increase proportionally with $\lambda$. The target distribution is physical only if $\lambda p\leq1$ (with the noiseless case $p=0$ understood separately). When preparation errors are also included, their probabilities are scaled independently according to the corresponding calibrated preparation model.

\paragraph*{Realizing the scaled instrument.}
A valid target probability distribution does not by itself imply that the instrument can be produced experimentally. In this protocol the available operations are the four \emph{already noisy} Pauli-dressed instruments $\mathcal B_{ab,k}$ introduced in Sec.~\ref{sec:pec}. We seek a sampling distribution $\boldsymbol\alpha(\lambda)$ satisfying, for every logical outcome $k$,
\begin{equation}
\overline{\mathcal M}_k^{(\lambda)}=\sum_{a,b}\alpha_{ab}(\lambda)\mathcal B_{ab,k},\qquad k\in\{0,1\}.
\end{equation}
The equality is required outcome by outcome because subsequent feedforward and reset operations depend on the logical record. Writing $\mathbf p=(p_{ab})_{ab}$ and $\mathbf p^{(\lambda)}=(p_{ab}^{(\lambda)})_{ab}$ in the order $(00,01,10,11)$, the convolution relation from Sec.~\ref{sec:pec} gives a relation $P\boldsymbol\alpha(\lambda)=\mathbf p^{(\lambda)}$. The target distribution is affine in $\lambda$: $\mathbf p^{(\lambda)}=(1-\lambda)\mathbf e_0+\lambda\mathbf p$. Since $P\boldsymbol\gamma=\mathbf e_0$ and $P\mathbf e_0=\mathbf p$, we immediately obtain
\begin{equation}
\begin{aligned}
\boldsymbol\alpha(\lambda) =P^{-1}\big[(1-\lambda)\mathbf e_0+\lambda\mathbf p\big] =(1-\lambda)\boldsymbol\gamma+\lambda\mathbf e_0.\label{eq:alpha}
\end{aligned}
\end{equation}
The same coefficients work for both measurement outcomes, so this is an identity of complete outcome-resolved instruments:
$\overline{\mathfrak M}^{(\lambda)}=\sum_{a,b}\alpha_{ab}(\lambda)\mathfrak B_{ab}$.

\begin{theorem}[Affine scaling and stochastic feasibility]\label{thm:feasibility}
Under the assumptions of Theorem~\ref{thm:convolution}, write $\boldsymbol\gamma=C^{-1}\mathbf e_0$ for the coefficient vector representing the ideal branch (equivalently $P^{-1}\mathbf e_0$ in the symmetric binary model). Let the target error distribution be $\mathbf p^{(\lambda)}=(1-\lambda)\mathbf e_0+\lambda\mathbf p$, where $\mathbf p=(p(g))_{g\in G}$, with $\lambda\geq1$, and retain the same dressed-operation convolution representation. Its unique coefficient vector is
\begin{equation}
\boldsymbol\alpha(\lambda)=(1-\lambda)\boldsymbol\gamma+\lambda\mathbf e_0.
\end{equation}
Within this fixed dressed-operation sampling protocol, exact stochastic realization is possible if and only if every component of $\boldsymbol\alpha(\lambda)$ is nonnegative. For $\lambda>1$ this is equivalent to $\gamma_g\leq0$ for each $g\ne0$ and $\lambda-(\lambda-1)\gamma_0\geq0$. In the prescribed nonidentity-error scaling, target physicality requires $\lambda[1-p(0)]\leq1$; this condition follows automatically from an exact nonnegative sampling representation but is useful to check independently.
\end{theorem}
\noindent\emph{Proof.} The native undressed branch corresponds to $e_0$, and its error distribution is $p$; the ideal branch corresponds to $\gamma$. Linearity gives the displayed affine identity. Because convolution preserves the sum of coefficients, $\sum_g\gamma_g=1$ and hence $\sum_g\alpha_g(\lambda)=1$. A nonnegative normalized coefficient vector is a probability distribution over accessible dressed instruments, and conversely the unique coefficient vector cannot be used as such a sampling distribution if any entry is negative. Resolving its identity and non-identity entries gives the inequalities. Finally, the prescribed scaling assigns probability $1-\lambda[1-p(0)]$ to the identity error and $\lambda p_g$ to each other error, proving the physicality condition. $\square$ \\

The operational scope of this criterion is essential. Nonvanishing Fourier eigenvalues determine the inverse in the selected convolution representation; they do not ensure that a prescribed amplified distribution is synthesizable with nonnegative weights. Conversely, physicality of the target error distribution alone does not ensure stochastic synthesis from the fixed dressed-operation family. The resulting boundary between algebraic cancellation and experimentally sampleable amplification is the principal PEC--ZNE distinction established here. It is not a no-go statement for additional controls, alternative instrument dressings, or physically equivalent decompositions of linearly dependent ideal maps.

The theorem states a necessary and sufficient criterion for exact sampling using these convolution coefficients. For the calibrated model, nonnegative sampling weights already imply that the target distribution is physical; the explicit physicality test is retained to distinguish target definition from protocol realizability.

Eq.~\eqref{eq:alpha} connects PEC and ZNE in this model. At $\lambda=0$, $\boldsymbol\alpha(0)=\boldsymbol\gamma$ reconstructs the ideal instrument by a signed combination; at $\lambda=1$, $\boldsymbol\alpha(1)=\mathbf e_0$ selects the native noisy instrument; and for $\lambda>1$ it represents a target with increased stochastic error probabilities. These coefficient vectors lie on an affine line in the common dressed-operation basis (Fig.~\ref{fig:PEC1}). This statement does \emph{not} equate the statistical estimators for PEC and ZNE: PEC samples signed weights to cancel errors directly, whereas ZNE first samples physically realizable noise strengths and only then extrapolates measured expectations.

The two feasibility conditions must be checked separately: (i) $\mathbf p^{(\lambda)}$ must be a probability distribution, and (ii) the coefficients $\boldsymbol\alpha(\lambda)$ must all be nonnegative so they can be used as probabilities for selecting dressed operations. Normalization of $\boldsymbol\alpha$ follows from Eq.~\eqref{eq:alpha}, because both $\boldsymbol\gamma$ and $\mathbf e_0$ sum to one. For $\lambda>1$, the entries are
\[
\alpha_{00}(\lambda)=\lambda-(\lambda-1)\gamma_{00}~~\mathrm{and}~~
\alpha_{ab}(\lambda)=-(\lambda-1)\gamma_{ab}  ~~\mathrm{for}~~(a,b)\ne(0,0).
\]
For an invertible calibrated model, exact stochastic sampling over these four operations is possible precisely when $\gamma_{ab}\leq0$ for $(a,b)\ne(0,0)$ and $\lambda- (\lambda-1)\gamma_{00}\geq0$, in addition to the target physicality condition $\lambda p\leq1$. When $\gamma_{00}>1$, the second inequality becomes $\lambda\leq\gamma_{00}/(\gamma_{00}-1)$. A positive off-identity PEC coefficient makes its corresponding $\alpha_{ab}(\lambda)$ negative for \emph{every} $\lambda>1$, even arbitrarily close to one. In that case Eq.~\eqref{eq:alpha} is still a valid signed linear identity, but it cannot be implemented as an exact random mixture of this dressed-operation basis. Appendix~\ref{app:alpha} examines asymmetric error profiles and an approximate alternative.

\paragraph*{Preparation-noise scaling.} The same construction applies to initial-state preparation, but its error probability $q$ must be calibrated separately. For the binary preparation distribution $\mathbf q=(1-q,q)^T$, define $Q_{cd}=q_{c\oplus d}$ and the PEC weights $\boldsymbol\eta=Q^{-1}\mathbf e_0$ as in Eq.~\eqref{eq:eta_coefficients}. The desired scaled distribution is $\mathbf q^{(\lambda)}=(1-\lambda)\mathbf e_0+\lambda\mathbf q$. Expressing the target state as a mixture of the two accessible dressed preparations gives
\begin{equation}
\label{eq:beta_affine}
\boldsymbol\beta(\lambda)=Q^{-1}\mathbf q^{(\lambda)}=(1-\lambda)\boldsymbol\eta+\lambda\mathbf e_0.
\end{equation}
This parallels Eq.~\eqref{eq:alpha}: $\boldsymbol\beta(0)=\boldsymbol\eta$ gives the signed preparation PEC inverse, $\boldsymbol\beta(1)=\mathbf e_0$ gives the native noisy preparation, and larger admissible $\lambda$ increases the stochastic preparation error. For $q\ne1/2$ the two coefficients are
\begin{eqnarray}
\overline\rho^{(\lambda)}&=& \beta_0(\lambda)\rho_0 + \beta_1(\lambda)\rho_1,~~\mathrm{where}~~
 \beta_0(\lambda) =\frac{1-(1+\lambda)q}{1-2q}~\mathrm{and}~ \beta_1(\lambda) =\frac{(\lambda-1)q}{1-2q}. \label{eq:beta_explicit}
\end{eqnarray}
The target is $\overline\rho^{(\lambda)}=(1-\lambda q)\rho_{\mathrm{id}}+\lambda q\mathcal X(\rho_{\mathrm{id}})$. In the assumed low-error regime $0<q<1/2$, the sampling weights are nonnegative for $1\leq\lambda\leq(1-q)/q$, a stricter requirement than $\lambda q\leq1$. Sampling the dressed preparation with $c=1$ means inserting an $X$ after RC preparation, with probability $\beta_1(\lambda)$ (Fig.~\ref{fig:PRC}). If $q=0$, the two preparations coincide with their ideal targets and $\boldsymbol\beta=(1,0)^T$ for all scales. Equations~\eqref{eq:alpha} and~\eqref{eq:beta_affine} instantiate the affine-scaling result of Theorem~\ref{thm:feasibility} for measurements and preparations, respectively; they do not require the two errors to share a calibration model.

\paragraph*{Extrapolation to zero noise.}
Once physically admissible scales have been chosen, the circuit is executed separately at each $\lambda\in\Lambda$ to estimate an observable $E(\lambda)$. If each of the finitely many preparation and measurement blocks depends affinely on the common $\lambda$, the expectation value is a polynomial in $\lambda$. This remains true for adaptive circuits: conditioning later gates on earlier measurement results produces a sum over outcome histories, and each history contributes a product of affine, outcome-conditioned maps. The conclusion assumes the calibrated model holds at every scale; drift, memory, leakage, or uncharacterized errors from the added dressing operations can spoil this dependence.

For distinct scales $\Lambda$, Richardson extrapolation estimates the zero-noise value using
\begin{equation}
\widehat E(0)=\sum_{\lambda\in\Lambda}w_\lambda\widehat E(\lambda),~~\mathrm{where}
~~
w_\lambda=\prod_{\substack{\lambda'\in\Lambda\\\lambda'\ne\lambda}}
\frac{-\lambda'}{\lambda-\lambda'}.
\end{equation}
These weights satisfy $\sum_\lambda w_\lambda=1$ and $\sum_\lambda w_\lambda\lambda^j=0$ for $1\leq j<|\Lambda|$. Thus, with $|\Lambda|$ scales, polynomial error terms through degree $|\Lambda|-1$ cancel; higher-order contributions generally remain~\cite{richardson1911approximate}. For instance, scales $\{1,2\}$ give $\widehat E(0)=2\widehat E(1)-\widehat E(2)$, which removes an error term linear in $\lambda$ but not a generic quadratic term.

\vspace{0.7em}
\noindent\textbf{Algorithm: ZNE for $N_{\rm P}$ Preparations and $N_{\rm M}$ Measurements}
\vspace{0.3em}\hrule\vspace{0.3em}
\textbf{Input:} Target circuit, characterized $p_{ab},q$, and candidate scale factors $\Lambda$.
\begin{enumerate}
\item For each $\lambda\in\Lambda$, verify both that the scaled preparation and measurement error distributions are physical and that $\boldsymbol\alpha(\lambda)$ and $\boldsymbol\beta(\lambda)$ are nonnegative and normalized.
\item For every preparation, implement RC (random $Z$) and insert $X$ with probability $\beta_1(\lambda)$.
\item For each MCM, implement measurement RC, sample $(a,b)$ with probability $\alpha_{ab}(\lambda)$, insert $X^b$ before and $X^a$ after the RC measurement, and track the corrected logical outcome for feedforward.
\item Execute the resulting circuits and estimate $E(\lambda)$ for each scale.
\item Combine the estimates with the Richardson weights to obtain $\widehat E(0)=\sum_{\lambda\in\Lambda}w_\lambda\widehat E(\lambda)$.
\end{enumerate}
\vspace{0.3em}\hrule\vspace{0.5em}
Finally, exact noise amplification does not imply variance-free extrapolation. The Richardson coefficients are generally signed, and for independent scale estimates the variance is
$\operatorname{Var}[\widehat E(0)]=\sum_{\lambda\in\Lambda}w_\lambda^2\operatorname{Var}[\widehat E(\lambda)]$.
If all scales have equal estimator variance, the variance is multiplied by $\sum_\lambda w_\lambda^2$ relative to one scale estimate; a fair resource comparison must also account for how the total shots are distributed among scales and for calibration cost~\cite{RevModPhys.95.045005}.

\section{Proof-of-principle demonstrations}

In this section, we present proof-of-principle demonstrations of the proposed quantum-instrument framework using iterative quantum phase estimation, an adaptive algorithm involving mid-circuit measurements, feedforward, and qubit reuse. We first perform numerical simulations using IBM Qiskit Aer under various measurement-error models to compare instrument-level PEC and ZNE with conventional readout correction. These simulations allow us to examine the effects of different measurement-error mechanisms and assess the performance of the proposed mitigation methods under controlled noise conditions. We then investigate the experimental implementation of the framework using a three-cycle adaptive phase-estimation circuit on a $20$-qubit superconducting quantum processor at KRISS. Using two qubits and experimentally calibrated measurement errors, we implement PEC and controlled stochastic noise amplification for ZNE within the same dressed-operation framework and compare their performance with conventional readout correction.

\label{sec:demo}
\begin{figure*}[t]
\includegraphics[width=0.98\textwidth]{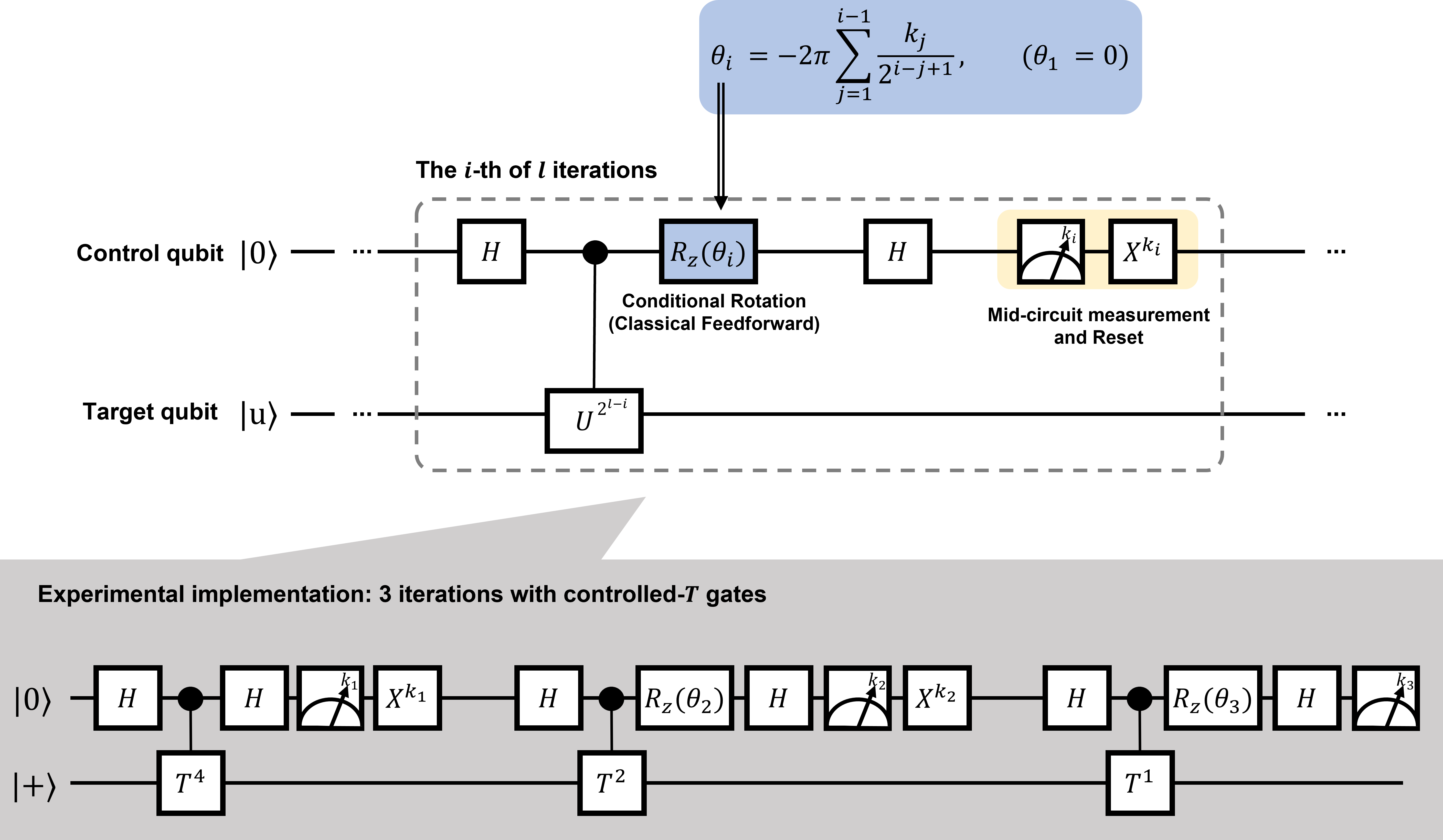}
 \caption{ (Upper) One iteration of adaptive one-qubit phase estimation (1QPE) is shown. In the $i$-th iteration, a controlled-$U^{2^{l-i}}$ gate acts on the target eigenstate $|u\rangle$, followed by the feedforward rotation $R_z(\theta_i)$, where $\theta_i=-2\pi\sum_{j=1}^{i-1} k_j/2^{i-j+1}$ and $\theta_1=0$. The measurement produces $k_i$, which controls the reset $X^{k_i}$ before reuse. The output bits find an estimate $\hat\varphi=\sum_{i=1}^{l}k_i/2^{l-i+1}$. (Lower) Experimental three-cycle 1QPE circuit ($l=3$) using controlled-$T$ gates ($T^4, T^2, T^1$) with the target qubit initialized in $\ket{+}$.}
  \label{fig:1QPE}
\end{figure*}

\begin{figure}[t]
\includegraphics[width=0.94\columnwidth]{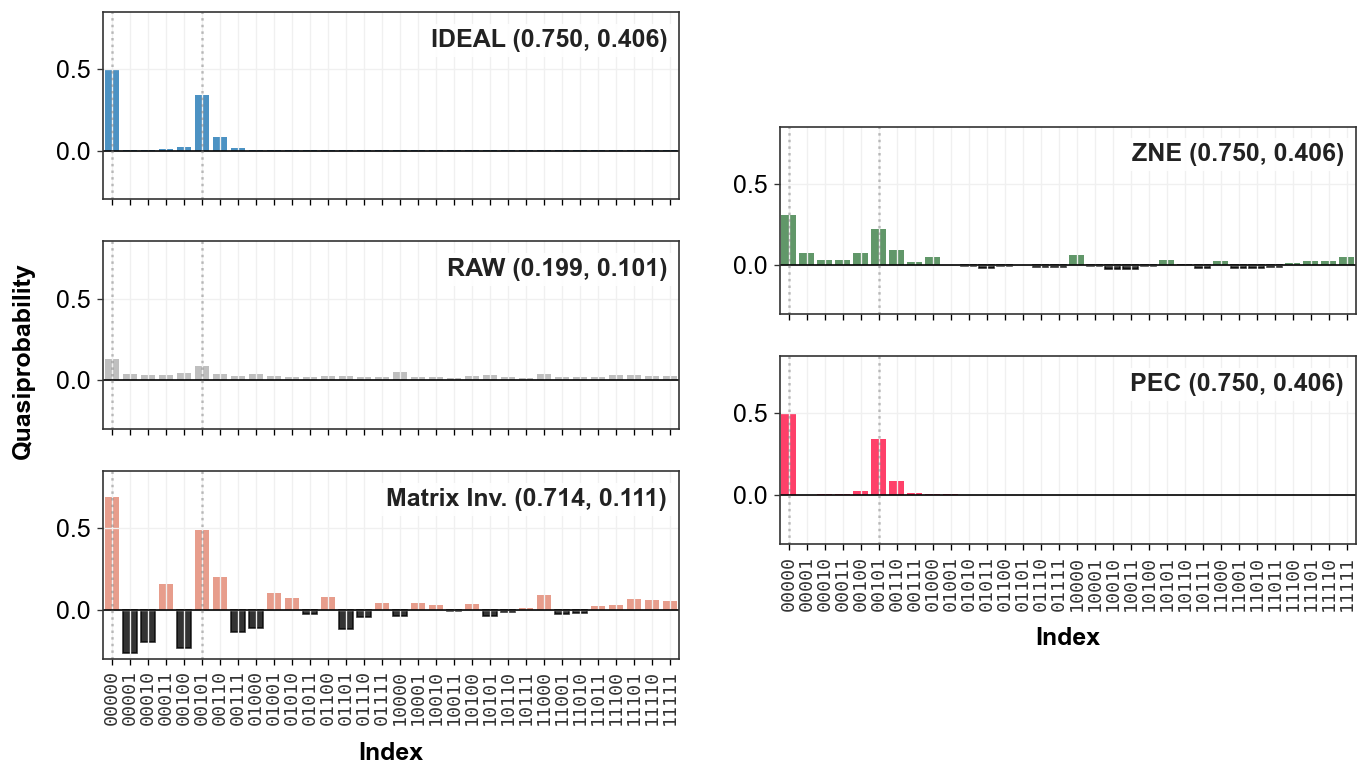}
    \caption{Exact-model phase-estimate distributions for five-cycle 1QPE are shown at total error probability $p=0.25$, error profile $(p_{01}:p_{10}:p_{11})=(0:0:1)$, and target eigenphases $\varphi\in\{0,1/6\}$. Panel labels report $(\langle\cos(2\pi\hat\varphi)\rangle,\langle\sin(2\pi\hat\varphi)\rangle)$. Black bars indicate negative reconstructed quasiprobabilities, not physical negative frequencies. In this example, second-order ZNE at $\lambda\in\{1,2,3\}$ recovers the displayed expectation coordinates but not necessarily the entire distribution; PEC reconstructs the ideal distribution within the model. Readout-only inversion does not mitigate all conditional-state errors.}
  \label{fig:sim1}
\end{figure}

\subsection{Numerical Demonstration of Instrument-Level Error Mitigation  }
\label{sec:simulation}

We investigate iterative one-qubit phase estimation (1QPE), as shown in Fig.~\ref{fig:1QPE}, in which an ancilla is repeatedly measured and reset, with subsequent operations conditioned on earlier measurement outcomes. We compare three error-mitigation approaches: conventional readout matrix inversion~\cite{PhysRevA.103.042605}, which corrects the effective readout error $p_{\mathrm{meas}}=p_{01}+p_{11}$; second-order ZNE using noise-scaling factors $\lambda\in\{1,2,3\}$; and instrument-level PEC. For ZNE, exact stochastic noise amplification is considered only when the sampling coefficients satisfy the nonnegativity conditions of Eq.~\eqref{eq:alpha}, which must be verified for each error profile.

For the evaluation, we consider the phase-expectation vector $\mathbf v=(\langle\cos(2\pi\hat\varphi)\rangle,\langle\sin(2\pi\hat\varphi)\rangle)$ and quantify its deviation from the ideal reference using the Euclidean distance $\Delta=\|\mathbf v-\mathbf v_{\mathrm{id}}\|_2$.
We perform both infinite-shot (exact-model) and finite-shot simulations with $30,000$ shots using a target qubit initialized to $\ket{u}=\ket{+}$ and an ancilla undergoing $l=5$ measurement-and-reset cycles. The controlled operations are generated by $U= \mathrm{diag} [1 ~~e^{i2\pi/6}]$ such that the target state contains eigenphases $\varphi=0$ and $\varphi=1/6$. Assuming ideal initial state preparation ($q=0$), we vary the total measurement-error probability over $p\in[0,0.25]$ and consider four error profiles specified by $(p_{01}:p_{10}:p_{11})$: a symmetric profile $(1:1:1)$ and three asymmetric profiles $(1:0:0)$, $(0:1:0)$, and $(0:0:1)$.

We first examine the phase-estimation probability distributions for the $(0:0:1)$ error profile at $p=0.25$, comparing the unmitigated results with classical readout matrix inversion, instrument-level PEC, and second-order ZNE, as shown in Fig.~\ref{fig:sim1}. For this profile, PEC reconstructs the ideal probability distribution within the assumed noise model, whereas matrix inversion leaves residual deviations. Second-order ZNE recovers the ideal phase-expectation coordinates in this case, although the reconstructed probability distribution does not necessarily coincide with the ideal distribution.

\begin{figure}[t]
\includegraphics[width=0.75\columnwidth]{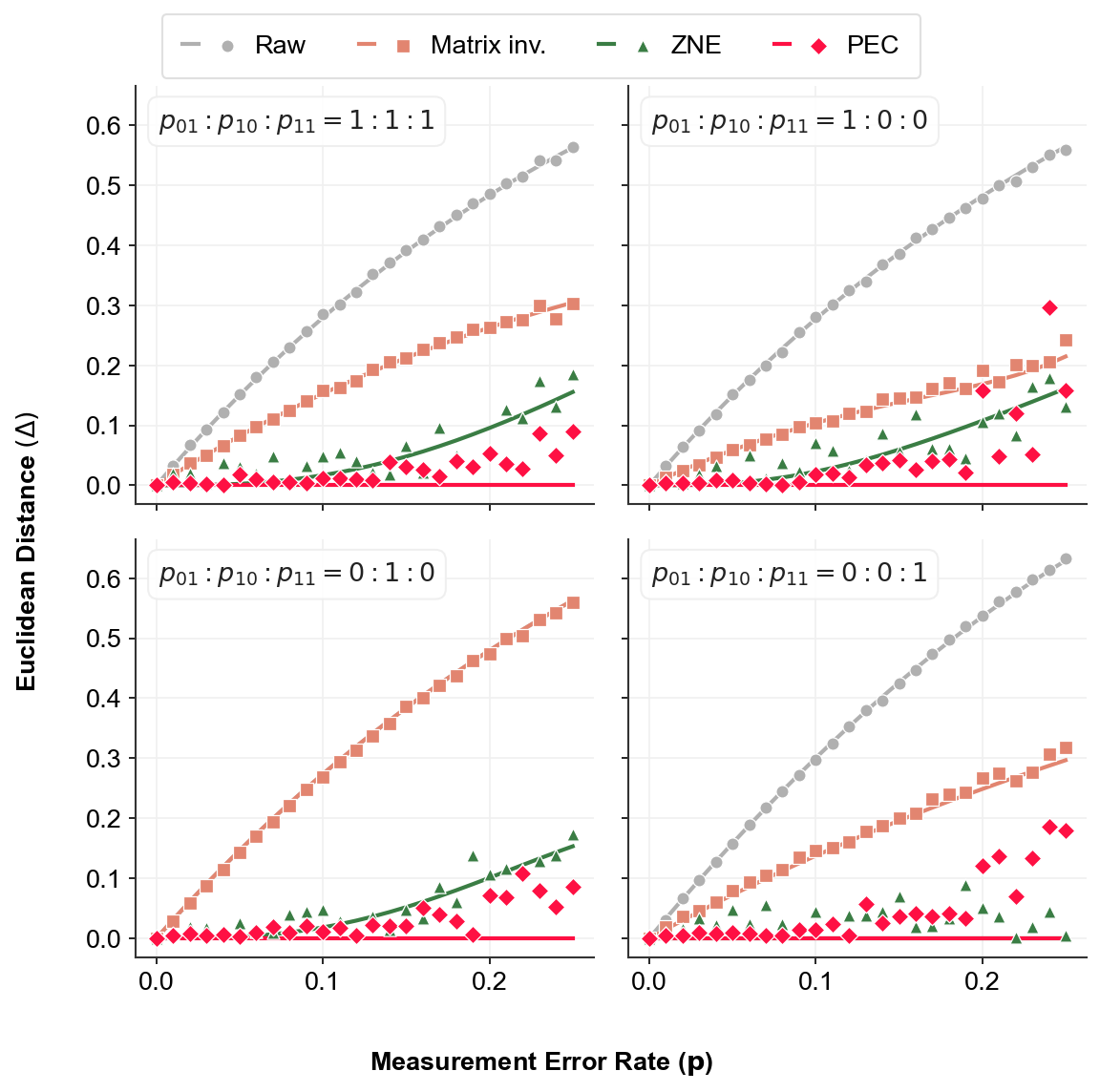}
  \caption{The comparison, phase-expectation error versus total measurement error probability $p$ in five-cycle 1QPE, is shown for four indicated error profiles $(p_{01}:p_{10}:p_{11})$. Solid curves show exact-model results; points represent 30,000-shot simulations. The metric $\Delta$ is the Euclidean distance from the ideal phase-expectation vector. For post-measurement-only errors (0:1:0), readout-matrix inversion has no effect because the outcome-assignment error is zero. PEC has zero analytical bias within the assumed model but exhibits finite-shot variability; finite-order ZNE can retain higher-order bias.}
  \label{fig:sim2}
\end{figure}

We further evaluate the mitigation performance across the four measurement-error profiles by calculating the Euclidean distance $\Delta$ between the estimated and ideal phase-expectation vectors as a function of $p$, as shown in Fig.~\ref{fig:sim2}. The unmitigated error generally increases with $p$, while the mitigation methods exhibit different behavior depending on the error profile. For the $(0:1:0)$ profile, where the measurement outcomes are ideal but the post-measurement states are disturbed, classical matrix inversion leaves the results unchanged. Instrument-level PEC removes the analytical bias within the assumed stochastic model, although finite-shot fluctuations remain. ZNE reduces the leading-order bias but can retain higher-order contributions due to finite-order extrapolation.

The numerical results have shown the feasibility of applying instrument-level PEC and ZNE to adaptive quantum circuits involving repeated mid-circuit measurements within the assumed stochastic noise model. Instrument-level PEC reconstructs the ideal outcome-conditioned operations, addressing errors in both classical measurement outcomes and post-measurement quantum states. ZNE provides a complementary approach to reducing measurement-induced bias through controlled noise amplification and extrapolation, subject to the physical sampling conditions.

\subsection{Experimental Demonstration on the KRISS Superconducting Quantum Processor}
\label{sec:exp}

We experimentally investigate the proposed quantum-instrument framework on a 20-qubit superconducting quantum processor developed at KRISS. The experiment uses two adjacent qubits, Q16 and Q17, with Q17 serving as the ancilla for mid-circuit measurements, measurement-conditioned feedforward, and qubit reuse. We implement a three-cycle iterative phase-estimation circuit using controlled-$T$ gates, with the target qubit initialized to $\ket{+}$ to probe eigenphases $\varphi=0$ and $\varphi=1/8$. The experiment enables the implementation of instrument-level PEC and stochastic noise amplification for ZNE using measurement errors characterized on the KRISS processor. Detailed device characteristics are provided in Appendix C.

Using 10,000 experimental shots (compared with 30,000 shots in the finite-shot simulations), we evaluate PEC, first-order ZNE ($\lambda \in \{1, 2\}$), and the matrix inversion baseline via the metric $\Delta$. Measurement error rates ($p_{ab}$) are inferred conditionally on ideal state preparation ($q=0$) and the reference-readout model (Appendix~\ref{app:characterize}). By repeating the characterization protocol $2^{14}$ times, we obtained the probabilities $p_{00} \approx 0.8879$, $p_{01} \approx 0.0245$, $p_{10} \approx 0.0226$, and $p_{11} \approx 0.0650$. Using these values, the quasi-probability coefficients for PEC were calculated as $\gamma_{00} \approx 1.1336$, $\gamma_{01} \approx -0.0274$, $\gamma_{10} \approx -0.0246$, and $\gamma_{11} \approx -0.0816$. For error scaling in ZNE at $\lambda = 2$, the corresponding exact sampling weights were determined to be $\alpha_{00}{(2)} \approx 0.8664$, $\alpha_{01}{(2)} \approx 0.0274$, $\alpha_{10}{(2)} \approx 0.0246$, and $\alpha_{11}{(2)} \approx 0.0816$. Finally, the matrix inversion baseline is applied by targeting the effective readout error, defined as $p_{\mathrm{meas}} = p_{01} + p_{11} \approx 0.0895$.

The PEC coefficients imply $\Gamma_{\rm M}\simeq1.2672$. For three independently mitigated MCMs and ideal preparation, the square of the absolute trajectory weight is $\Gamma_{\rm M}^{6}\simeq4.14$, giving a worst-case second-moment bound relative to a unit-bounded raw outcome. This is not a measured variance ratio and omits uncertainty in the calibrated $p_{ab}$ and additional gate errors. First-order ZNE with scales $\{1,2\}$ uses Richardson coefficients $(2,-1)$: for independent scale estimates, its variance is $4\operatorname{Var}(\widehat E_1)+\operatorname{Var}(\widehat E_2)$. Total costs also depend on allocation of shots between these scales and calibration.

\begin{figure}[t]
\includegraphics[width=0.84\columnwidth]{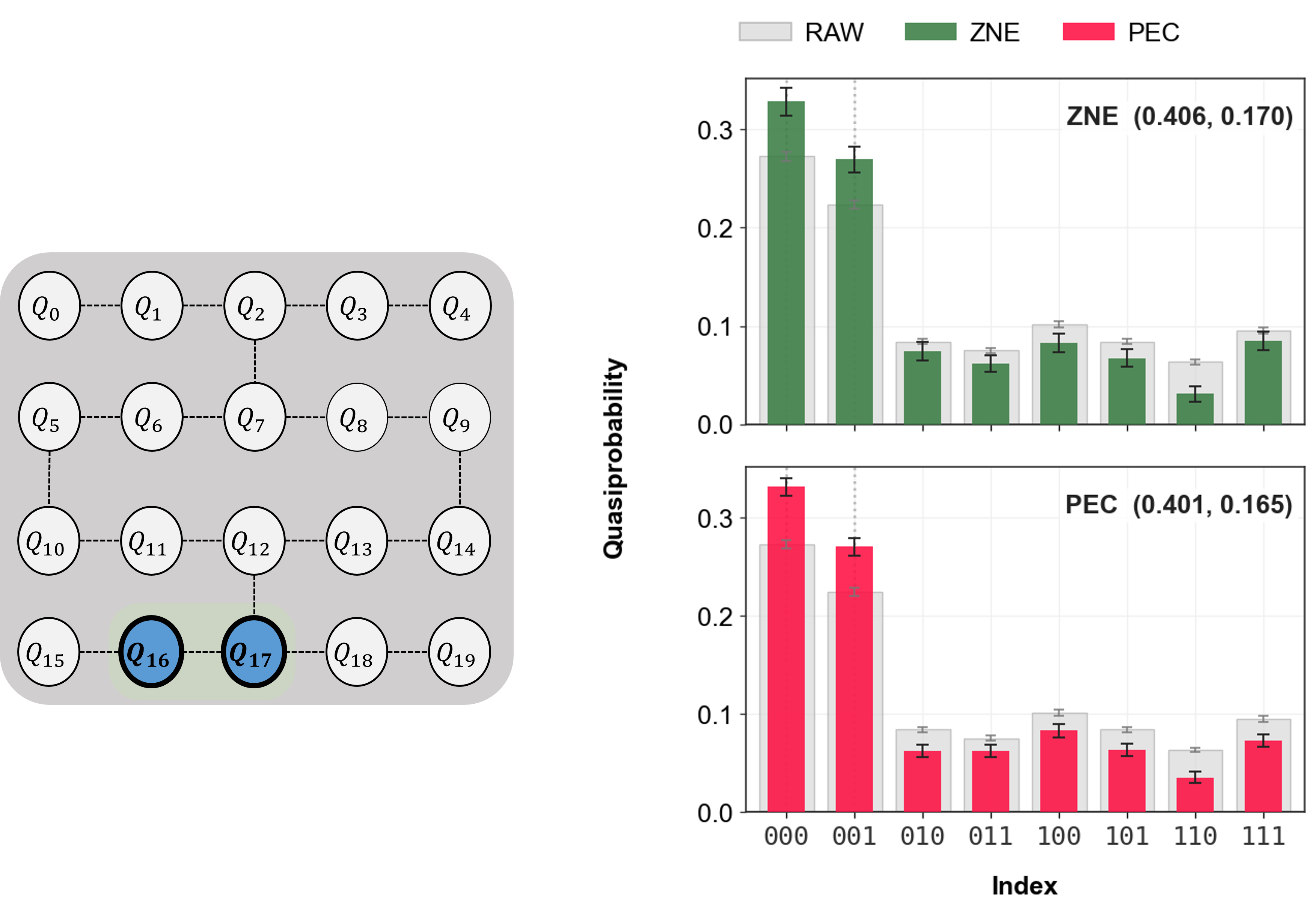}
    \caption{ (Left) Layout of the 20-qubit superconducting quantum processor developed at KRISS, with Q16 and Q17 used in the experiment. (Right) Experimental phase-estimation outcome distributions for the three-cycle 1QPE circuit. Gray bars represent unmitigated results, while green and red bars represent first-order ZNE with $\lambda\in\{1,2\}$ and instrument-level PEC, respectively. The ideal distribution assigns probability $1/2$ to each of the outcomes 000 and 001. Panel annotations indicate the phase-expectation vectors $(\langle\cos(2\pi\hat{\varphi})\rangle,\langle\sin(2\pi\hat{\varphi})\rangle)$. Error bars represent standard deviations of the estimated outcome probabilities. }
  \label{fig:exp}
\end{figure}

\begin{figure}[t]
\includegraphics[width=0.84\columnwidth]{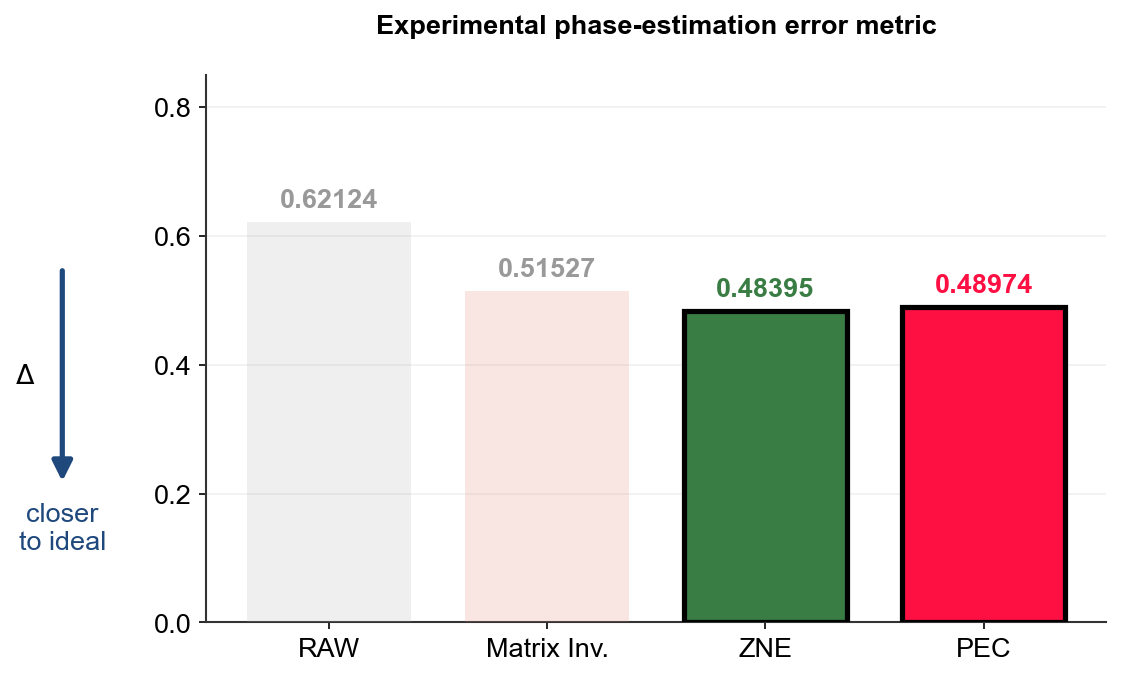}
\caption{Euclidean distance $\Delta$ between experimental and ideal phase-expectation vectors for the unmitigated circuit, readout-matrix inversion, first-order ZNE, and PEC. Bars are the point estimates in Table~\ref{tab:qpe_performance_stats}; smaller values indicate closer agreement with the ideal.}
\label{fig:experimental_metric}
\end{figure}

Figure~\ref{fig:exp} shows the processor layout and the experimental phase-estimation outcome distributions for the unmitigated, first-order ZNE, and PEC circuits. Figure~\ref{fig:experimental_metric} and Table~\ref{tab:qpe_performance_stats} report the point estimates using the Euclidean distance $\Delta$ to the ideal phase expectation. Compared to the unmitigated raw data ($\Delta \approx 0.621$), ZNE ($\Delta \approx 0.484$) and PEC ($\Delta \approx 0.490$) reduce the distance by 22.1\% and 21.2\%, respectively, giving lower point estimates than the matrix-inversion baseline ($\Delta \approx 0.515$, 17.1\% reduction). Residual deviations remain. The differences relative to matrix inversion are modest in absolute units ($\Delta_{\mathrm{MI}}-\Delta_{\mathrm{ZNE}}\simeq0.0313$ and $\Delta_{\mathrm{MI}}-\Delta_{\mathrm{PEC}}\simeq0.0255$).

\begin{table}[t]
\small
\caption{Experimental phase-expectation vectors $(\langle\cos(2\pi\hat\varphi)\rangle,\langle\sin(2\pi\hat\varphi)\rangle)$ and Euclidean distances $\Delta$ from the ideal reference. The ideal row assumes error-free preparation, gates, and measurements. Reported values represent point estimates without uncertainty intervals.}
\label{tab:qpe_performance_stats}
\setlength{\tabcolsep}{8pt}
\begin{tabular}{l c c}
\hline \hline
\textbf{Method} & $(\langle \cos(2\pi\hat{\varphi}) \rangle, \langle \sin(2\pi\hat{\varphi}) \rangle)$ & $\Delta$ \\
\hline
IDEAL       & $(0.85355,\, 0.35355)$ & -- \\
RAW         & $(0.28390,\, 0.10569)$ & $0.62124$ \\
Matrix Inv  & $(0.40077,\, 0.10761)$ & $0.51527$ \\
\textbf{ZNE} & ${(0.40561,\, 0.17040)}$ & $\mathbf{0.48395}$ \\
\textbf{PEC} & ${(0.40148,\, 0.16521)}$ & $\mathbf{0.48974}$ \\
\hline \hline
\end{tabular}
\end{table}

\section{Conclusion}

In summary, we have developed a unified framework for probabilistic error cancellation (PEC) and zero-noise extrapolation (ZNE) of mid-circuit quantum measurements. Using an outcome-resolved quantum-instrument description, we formulate ideal measurement reconstruction and controlled noise amplification using the same Pauli-dressed operations. We establish an affine relation between the PEC and noise-scaling coefficients and identify the conditions for physically realizable stochastic noise amplification. We investigate the framework through numerical simulations of iterative quantum phase estimation under different measurement-error models and a proof-of-principle experiment on a $20$-qubit superconducting quantum processor at the Korea Research Institute of Standards and Science (KRISS). In the experiment, both PEC and ZNE produce phase-expectation point estimates closer to the ideal reference than conventional readout correction.

Our results show that errors affecting classical measurement outcomes and conditional post-measurement quantum states can be addressed within an outcome-resolved quantum-instrument framework. The relationship between PEC reconstruction and stochastic noise scaling provides a systematic way to formulate both mitigation methods using the same experimentally accessible operations. The derived nonnegativity conditions determine when the prescribed noise amplification can be implemented through probabilistic sampling, clarifying the operational distinction between PEC and ZNE.

The proposed framework provides a basis for mitigating measurement-induced errors in adaptive quantum circuits involving feedforward and qubit reuse. Future work may extend the approach to multiqubit measurements and more general noise models, including leakage and temporal correlations. Further investigations of alternative dressed-operation protocols, sampling overhead, and experimental performance may help assess its applicability to larger-scale adaptive quantum computations.

\section*{Data and Software Availability }
The data and software supporting the findings of this study are publicly available at \url{https://github.com/Hyunsung619/mcm_data}. The repository contains the source code, analysis scripts, and relevant data required to reproduce the results presented in this article.

\section*{Acknowledgements}
This work is supported by the Institute for Information \& Communication Technology Promotion (IITP) (RS-2025-02304540, RS-2025-25464876, RS-2025-25464616).
\bibliographystyle{unsrt}
\bibliography{ref}

@article{PhysRevLett.133.150602,
  author = {Elisa Bäumer and Vinay Tripathi and Alireza Seif and Daniel Lidar and Derek S. Wang},
  title = {{Quantum Fourier Transform Using Dynamic Circuits}},
  year = {2024},
  doi = {10.1103/PhysRevLett.133.150602},
  journal = {Physical Review Letters},
  volume = {133},
  eid = {150602}
}

@article{RevModPhys.87.307,
  author = {Barbara M. Terhal},
  title = {{Quantum Error Correction for Quantum Memories}},
  year = {2015},
  doi = {10.1103/RevModPhys.87.307},
  journal = {Reviews of Modern Physics},
  volume = {87},
  pages = {307--346}
}

@inproceedings{548464,
  author = {Peter W. Shor},
  title = {{Fault-Tolerant Quantum Computation}},
  year = {1996},
  doi = {10.1109/SFCS.1996.548464},
  booktitle = {Proceedings of the 37th Annual Symposium on Foundations of Computer Science},
  pages = {56--65}
}

@article{PhysRevLett.76.3228,
  author = {Robert B. Griffiths and Chi-Sheng Niu},
  title = {{Semiclassical Fourier Transform for Quantum Computation}},
  year = {1996},
  doi = {10.1103/PhysRevLett.76.3228},
  journal = {Physical Review Letters},
  volume = {76},
  pages = {3228--3231}
}

@article{PhysRevA.103.042605,
  author = {Sergey Bravyi and Sarah Sheldon and Abhinav Kandala and David C. McKay and Jay M. Gambetta},
  title = {{Mitigating Measurement Errors in Multiqubit Experiments}},
  year = {2021},
  doi = {10.1103/PhysRevA.103.042605},
  journal = {Physical Review A},
  volume = {103},
  eid = {042605}
}

@article{PRXQuantum.2.040326,
  author = {Paul D. Nation and Hwajung Kang and Neereja Sundaresan and Jay M. Gambetta},
  title = {{Scalable Mitigation of Measurement Errors on Quantum Computers}},
  year = {2021},
  doi = {10.1103/PRXQuantum.2.040326},
  journal = {PRX Quantum},
  volume = {2},
  eid = {040326}
}

@article{PhysRevA.105.032620,
  author = {Ewout van den Berg and Zlatko K. Minev and Kristan Temme},
  title = {{Model-Free Readout-Error Mitigation for Quantum Expectation Values}},
  year = {2022},
  doi = {10.1103/PhysRevA.105.032620},
  journal = {Physical Review A},
  volume = {105},
  eid = {032620}
}

@article{Maciejewski2019,
  author = {Filip B. Maciejewski and Zoltán Zimborás and Michał Oszmaniec},
  title = {{Mitigation of Readout Noise in Near-Term Quantum Devices by Classical Post-Processing Based on Detector Tomography}},
  year = {2020},
  doi = {10.22331/q-2020-04-24-257},
  journal = {Quantum},
  volume = {4},
  eid = {257}
}

@article{9142431,
  author = {Hyeokjea Kwon and Joonwoo Bae},
  title = {{A Hybrid Quantum-Classical Approach to Mitigating Measurement Errors in Quantum Algorithms}},
  year = {2021},
  doi = {10.1109/TC.2020.3017888},
  journal = {IEEE Transactions on Computers},
  volume = {70},
  pages = {1401--1411}
}

@article{PhysRevLett.119.180509,
  author = {Kristan Temme and Sergey Bravyi and Jay M. Gambetta},
  title = {{Error Mitigation for Short-Depth Quantum Circuits}},
  year = {2017},
  doi = {10.1103/PhysRevLett.119.180509},
  journal = {Physical Review Letters},
  volume = {119},
  eid = {180509}
}

@article{PhysRevX.8.031027,
  author = {Suguru Endo and Simon C. Benjamin and Ying Li},
  title = {{Practical Quantum Error Mitigation for Near-Future Applications}},
  year = {2018},
  doi = {10.1103/PhysRevX.8.031027},
  journal = {Physical Review X},
  volume = {8},
  eid = {031027}
}

@article{PhysRevA.109.062617,
  author = {Riddhi S. Gupta and Ewout van den Berg and Maika Takita and Diego Ristè and Kristan Temme and Abhinav Kandala},
  title = {{Probabilistic Error Cancellation for Dynamic Quantum Circuits}},
  year = {2024},
  doi = {10.1103/PhysRevA.109.062617},
  journal = {Physical Review A},
  volume = {109},
  eid = {062617}
}

@article{PRXQuantum.6.010307,
  author = {Akel Hashim and Arnaud Carignan-Dugas and Larry Chen and Christian Jünger and Neelay Fruitwala and Yilun Xu and Gang Huang and Joel J. Wallman and Irfan Siddiqi},
  title = {{Quasiprobabilistic Readout Correction of Midcircuit Measurements for Adaptive Feedback via Measurement Randomized Compiling}},
  year = {2025},
  doi = {10.1103/PRXQuantum.6.010307},
  journal = {PRX Quantum},
  volume = {6},
  eid = {010307}
}

@article{cj89-4h5t,
  author = {Jin Ming Koh and Dax Enshan Koh and Jayne Thompson},
  title = {{Readout Error Mitigation for Mid-Circuit Measurements and Feedforward}},
  year = {2026},
  doi = {10.1103/cj89-4h5t},
  journal = {PRX Quantum},
  volume = {7},
  eid = {010317}
}

@article{PhysRevX.7.021050,
  author = {Ying Li and Simon C. Benjamin},
  title = {{Efficient Variational Quantum Simulator Incorporating Active Error Minimization}},
  year = {2017},
  doi = {10.1103/PhysRevX.7.021050},
  journal = {Physical Review X},
  volume = {7},
  eid = {021050}
}

@article{PhysRevA.94.052325,
  author = {Joel J. Wallman and Joseph Emerson},
  title = {{Noise Tailoring for Scalable Quantum Computation via Randomized Compiling}},
  year = {2016},
  doi = {10.1103/PhysRevA.94.052325},
  journal = {Physical Review A},
  volume = {94},
  eid = {052325}
}

@article{RevModPhys.95.045005,
  author = {Zhenyu Cai and Ryan Babbush and Simon C. Benjamin and Suguru Endo and William J. Huggins and Ying Li and Jarrod R. McClean and Thomas E. O’Brien},
  title = {{Quantum Error Mitigation}},
  year = {2023},
  doi = {10.1103/RevModPhys.95.045005},
  journal = {Reviews of Modern Physics},
  volume = {95},
  eid = {045005}
}

@article{HinesProctor2025,
  author = {Jordan Hines and Timothy Proctor},
  title = {{Pauli Noise Learning for Mid-Circuit Measurements}},
  year = {2025},
  doi = {10.1103/PhysRevLett.134.020602},
  journal = {Physical Review Letters},
  volume = {134},
  eid = {020602}
}

@article{ZhangChenLiuJiang2025,
  author = {Zhihan Zhang and Senrui Chen and Yunchao Liu and Liang Jiang},
  title = {{Generalized Cycle Benchmarking Algorithm for Characterizing Midcircuit Measurements}},
  year = {2025},
  doi = {10.1103/PRXQuantum.6.010310},
  journal = {PRX Quantum},
  volume = {6},
  eid = {010310}
}

@misc{beale2023randomized,
  author = {Stefanie J. Beale and Joel J. Wallman},
  title = {{Randomized Compiling for Subsystem Measurements}},
  year = {2023},
  eprint = {2304.06599},
  archivePrefix = {arXiv},
  primaryClass = {quant-ph},
  url = {https://arxiv.org/abs/2304.06599}
}

@article{richardson1911approximate,
  author = {Lewis Fry Richardson},
  title = {{The Approximate Arithmetical Solution by Finite Differences of Physical Problems Involving Differential Equations, with an Application to the Stresses in a Masonry Dam}},
  journal = {Philosophical Transactions of the Royal Society of London. Series A},
  volume = {210},
  pages = {307--357},
  year = {1911}
}

\clearpage

\appendix

\section{Characterization of the Measurement Error Rates} \label{app:characterize}

Within the computational-subspace, binary-outcome and randomized-compiling assumptions of Proposition~\ref{prop:tailoring}, the averaged instrument $\overline{\mathfrak M}=\{\overline{\mathcal M}_k\}_{k=0,1}$ has four stochastic error weights $p_{ab}$ shared by its outcome-conditioned operations. These weights specify the effective instrument; they should not, without further modeling, be interpreted as a unique microscopic decomposition of detector error and measurement backaction. The corresponding flagged channel, which explicitly stores the classical outcome, is
\begin{equation}
    \widehat{\overline{\mathcal M}}(\cdot) = \sum_{a,b,k \in \{0,1\}}p_{ab}[\mathcal X^a\circ\mathcal M_k\circ\mathcal X^b](\cdot)\otimes|k\rangle_C\langle k|.
\end{equation}

For an exactly prepared input $|0\rangle\langle 0|$, applying this flagged instrument channel yields:
\begin{equation}
\begin{aligned}
    \widehat{\overline{\mathcal M}}(|0\rangle\langle 0|) &= \left( p_{00}|0\rangle\langle 0| + p_{10}|1\rangle\langle 1| \right) \otimes |0\rangle_C\langle 0| \\
    &\quad + \left( p_{11}|0\rangle\langle 0| + p_{01}|1\rangle\langle 1| \right) \otimes |1\rangle_C\langle 1|.
\end{aligned}
\end{equation}
For outcomes with nonzero probabilities $s_k$, the conditional post-measurement states $\sigma_k$ are:
\begin{equation}
\begin{aligned}
    \sigma_0 &= \frac{p_{00}}{p_{00}+p_{10}}|0\rangle\langle 0| + \frac{p_{10}}{p_{00}+p_{10}}|1\rangle\langle 1|  \\~~\mathrm{and}
~~
    \sigma_1 &= \frac{p_{11}}{p_{01}+p_{11}}|0\rangle\langle 0| + \frac{p_{01}}{p_{01}+p_{11}}|1\rangle\langle 1|.
\end{aligned}
\end{equation}

Writing $s_k=\Pr(k)$ and $r_{j|k}=\langle j|\sigma_k|j\rangle$,
the joint output probabilities give the explicit extraction identities
\begin{equation}
 p_{00}=s_0r_{0|0},\quad p_{10}=s_0r_{1|0},\quad
 p_{11}=s_1r_{0|1}~~\mathrm{and}
~~ p_{01}=s_1r_{1|1}.
\end{equation}
The following procedure infers these effective coefficients from corrected joint frequencies under explicit trusted-input and reference-readout assumptions:

\vspace{0.5em}
\noindent \textbf{1. State Preparation ($|0\rangle$):}
The qubit is initialized to $|0\rangle$ via passive reset (idling much longer than the relaxation time $T_1$). The inference below assumes a trusted $|0\rangle$ input; residual reset error must be calibrated or propagated as a systematic uncertainty.

\noindent \textbf{2. First Measurement (RC):}
The randomized-compiled measurement yields the corrected logical outcome $k=k_{\mathrm{phys}}\oplus x$; this corrected bit, rather than the raw detector bit, must be retained in the joint frequencies. Independent $Z$ twirls are omitted in this particular population-characterization sequence because the prepared input and reference measurement are both diagonal in the computational basis. This omission does not establish the full instrument-twirling reduction for arbitrary coherent inputs and downstream operations.

\noindent \textbf{3. Second Measurement (Z-basis):}
A $Z$-basis reference measurement is performed on $\sigma_k$. Its calibrated classical confusion matrix is inverted to estimate the output populations, conditional on the matrix being valid and sufficiently well conditioned. This correction addresses the reference measurement record; it does not certify the reference measurement as noiseless or remove quantum-state disturbances.

\noindent \textbf{4. Parameter Extraction:}
After correcting and independently validating the second readout, multiplying the estimated populations of $\sigma_0$ and $\sigma_1$ by their respective marginal probabilities ($p_{00} + p_{10}$ for $k=0$; $p_{01} + p_{11}$ for $k=1$) yields all four $p_{ab}$ parameters. Practically, since $k=1$ indicates a rare baseline measurement error, accurately estimating $\sigma_1$ requires a large number of experimental shots. The extracted parameters therefore characterize the calibrated RC-averaged measurement model conditional on these assumptions. Uncertainty intervals should propagate finite counts, calibration-matrix uncertainty, residual reset errors, and drift. This calibration is the input to the group-convolution PEC inverse and the exact ZNE sampling test, so its conditioning directly affects both procedures.

\section{Analytical Conditions for Negative Weights in ZNE Error Scaling} \label{app:alpha}

This appendix refines the fixed-basis feasibility criterion of Theorem~\ref{thm:feasibility} and distinguishes exact from approximate amplification. In the calibrated convolution representation, the exact sampling weights $\alpha_{ab}(\lambda)$ for $\lambda > 1$ are determined by the PEC coefficients $\gamma_{ab}$ through Eq.~\eqref{eq:alpha}. To analyze the non-negativity condition $\alpha_{ab}(\lambda) \ge 0$, we expand $\gamma_{ab}$ in terms of the total measurement error probability $p = p_{01} + p_{10} + p_{11}$. For a perturbative discussion in the low-error regime, we retain terms through second order in $p$; neglected $\mathcal{O}(p^3)$ contributions need not be negligible for every device or error profile. Using $p_{00} = 1 - p$, applying the Taylor expansion $(1-x)^{-1} \approx 1 + x + x^2$ to the denominators in Eq.~\eqref{eq:gamma_coefficients} yields:

\begin{align}
\boldsymbol\alpha(\lambda) &=
(1-\lambda) \boldsymbol\gamma +
\lambda\mathbf e_0 \nonumber \\
&\approx (1-\lambda) \begin{pmatrix}
1 + p + p^2 + p_{01}^2 + p_{10}^2 + p_{11}^2 \\
-p_{01} + 2(p_{10}p_{11} - p_{01}p) \\
-p_{10} + 2(p_{01}p_{11} - p_{10}p) \\
-p_{11} + 2(p_{01}p_{10} - p_{11}p)
\end{pmatrix} + \lambda\mathbf e_0.
\end{align}

Independently of the perturbative expansion, for $\gamma_{00}>1$ the exact condition $\alpha_{00}(\lambda) \ge 0$ imposes $\lambda \le \gamma_{00}/(\gamma_{00}-1)$. If $\gamma_{00}=1$, this component imposes no upper bound on $\lambda>1$; for $\gamma_{00}<1$, the non-negativity inequality must be treated separately. For $\gamma_{00}>1$, a scale factor above the upper bound would give $\alpha_{00}(\lambda)<0$, so the proposed exact probabilistic sampling scheme would not apply.

For $ab \neq 00$ and $\lambda>1$, since $(1-\lambda) < 0$, the condition $\alpha_{ab}(\lambda) \ge 0$ is equivalent to $\gamma_{ab} \le 0$. At sufficiently low noise, the negative first-order term dominates when the corresponding error component is nonzero. For sufficiently asymmetric profiles, second-order cross terms can yield positive off-identity PEC coefficients and hence negative exact ZNE sampling weights. The perturbative expression is illustrative only: feasibility must be decided using the exact coefficients, not this truncated expansion.

When the exact coefficients are not nonnegative, one may adopt the following approximate scaling scheme. It samples a different physical mixture and must not be identified with the exact affine target or used to claim exact Richardson cancellation. This approach directly mixes the basis channels, ensuring physical sampling probabilities under the condition $(\lambda - 1)p \le 1$ while introducing an $\mathcal{O}(p^2)$ model deviation for fixed $\lambda$ in the low-noise regime:
\begin{align}
    \mathcal M_{k,\mathrm{approx}}^{(\lambda)}(\cdot) &= (1-(\lambda-1)p)\mathcal B_{00,k}(\cdot) + (\lambda-1)\sum_{ab \neq 00} p_{ab} \mathcal B_{ab,k}(\cdot) \nonumber \\
    &\approx (1 - \lambda p)\mathcal M_k(\cdot) + \lambda p \mathcal N_k(\cdot).
\end{align}

\paragraph*{Quantified approximation error.}
The $\mathcal O(p^2)$ statement concerns one scaled instrument at fixed $\lambda$, not an arbitrary composed adaptive circuit. Write the error distribution on the binary error-label group as $\mathbf p=e_0+s$, where $s_0=-p$, $s_g=p_g$ for $g\ne0$, and $\|s\|_1=2p$. The approximate mixture has error distribution
\begin{equation}
\mathbf p_{\mathrm{approx}}^{(\lambda)}
=\mathbf p+(\lambda-1)(s*\mathbf p)
=e_0+\lambda s+(\lambda-1)(s*s),
\end{equation}
where $*$ denotes group convolution. The intended affine target is $e_0+\lambda s$, so Young's inequality on the finite group yields
\begin{equation}\label{eq:approx_bound}
\|\mathbf p_{\mathrm{approx}}^{(\lambda)}-\mathbf p^{(\lambda)}\|_1
\leq(\lambda-1)\|s\|_1^2=4(\lambda-1)p^2.
\end{equation}
Because each ideal, outcome-flagged error-label operation is a completely positive trace-preserving channel, the triangle inequality bounds the diamond-norm distance between the two flagged channels by this coefficient $\ell_1$ bound. This does not assert the same bound for individually normalized conditional output states, whose normalization can amplify differences for rare outcomes. This is a bound on the calibrated stochastic model, not on unmodeled gate errors, leakage, or drift. For a circuit containing multiple approximate scaled blocks, the total deviation requires an additional composition bound; it is not supplied by Eq.~\eqref{eq:approx_bound} alone. Nor is the displayed $\mathcal O(p^2)$ estimate uniform for scale factors growing with $1/p$.

\section{Device Characteristics} \label{app:calibration}

For completeness, Table~\ref{tab:device_params} reports the hardware parameters relevant to the proof-of-principle demonstration in Sec.~III. Detailed specifications of the 2-qubit subsystem evaluated on the KRISS 20-qubit superconducting quantum processor are summarized here. The 1-qubit phase estimation experiment was executed using adjacent qubits Q16 and Q17, with Q17 serving as the control qubit. The reported calibration parameters were measured immediately prior to the experiment.

\begin{table}[h]
\caption{Device parameters for the two-qubit subsystem of the KRISS 20-qubit superconducting processor. $\omega_{01}$ and $\omega_r$ denote the qubit and readout-resonator angular frequencies; $T_1$ and $T_2$ are the energy-relaxation and Hahn-echo coherence times. Gate fidelities were assessed by randomized benchmarking.}
\label{tab:device_params}
\renewcommand{\arraystretch}{1.3}
\begin{tabular}{l|c|c}
\hline \hline
\textbf{1Q Parameter} & \makebox[2.4cm][c]{Q16} & \makebox[2.4cm][c]{Q17} \\ \hline
Qubit Freq. ($\omega_{01}/2\pi$) [GHz] & 5.113 & 5.172 \\
Readout Freq. ($\omega_{r}/2\pi$) [GHz] & 6.988 & 7.069 \\ \hline
$T_1$ [$\mu\text{s}$] & 38.2 & 11.6 \\
$T_2$ Echo [$\mu\text{s}$] & 47.7 & 52.7 \\ \hline
1Q Gate Fidelity & 0.999 & 0.995 \\
1Q Gate Duration [ns] & 64 & 64 \\ \hline \hline
\textbf{2Q Parameter} & \multicolumn{2}{c}{Q16--Q17} \\ \hline
2Q Gate Fidelity & \multicolumn{2}{c}{0.931} \\
2Q Gate Duration [ns] & \multicolumn{2}{c}{560} \\
\hline \hline
\end{tabular}
\end{table}

\end{document}